\documentclass[12pt]{article}

\usepackage[english]{babel}
\usepackage[utf8x]{inputenc}
\usepackage[T1]{fontenc}

\usepackage{amsmath}
\usepackage{physics}
\usepackage{float}
\usepackage{graphicx}
\usepackage[colorinlistoftodos]{todonotes}
\usepackage[colorlinks=true, allcolors=blue]{hyperref}
\usepackage{xfrac}
\usepackage{units}
\usepackage{amssymb}
\usepackage{dsfont}
\usepackage{epstopdf}
\usepackage{subfig}
\usepackage{tikz}
\usetikzlibrary{calc}
\usepackage{nicefrac}
\usepackage{mathtools}
\usepackage{bm}
\usepackage{epstopdf}
\usepackage[title]{appendix}
\usepackage[affil-it]{authblk}

\DeclarePairedDelimiter{\floor}{\lfloor}{\rfloor}

\providecommand{\keywords}[1]{\textbf{\textit{Keywords---}} #1}

\title{Non-linear ladder operators and coherent states for the 2:1 oscillator}
\date{}

\author[1,2]{James Moran\thanks{Corresponding Author: james.moran@umontreal.ca}}

\author[2,3]{V\'eronique Hussin\thanks{hussin@dms.umontreal.ca}}

\author[4]{Ian Marquette\thanks{i.marquette@uq.edu.au}}

\affil[1]{D\'epartement de physique, Universit\'e de Montr\'eal, Montr\'eal, Qu\'ebec, H3C 3J7, Canada}
\affil[2]{Centre de recherches math\'ematiques, Universit\'e de Montr\'eal, Montr\'eal, Qu\'ebec, H3C 3J7, Canada}
\affil[3]{D\'epartement de math\'ematiques et de statistique, Universit\'e de Montr\'eal, Montr\'eal, Qu\'ebec, H3C 3J7, Canada}
\affil[4]{School of Mathematics and Physics, The University of Queensland, Brisbane, QLD 4072, Australia}
\begin{document}
\maketitle
\begin{abstract}
The 2:1 two-dimensional anisotropic quantum harmonic oscillator is considered and new sets of states are defined by means of normal-ordering non-linear operators through the use of non-commutative binomial theorems as well as solving recurrence relations. The states generated are good candidates for the natural generalisation of the $\mathfrak{su}(2)$ coherent states of the two-dimensional isotropic oscillator. The two-dimensional non-linear generalised ladder operators lead to several chains of states which are connected in a non trivial way. The uncertainty relations of the defining chain of states are calculated and it is found that they admit a resolution of the identity and the spatial distribution of the wavefunction produces Lissajous figures in correspondence with the classical 2:1 oscillator. 
\end{abstract}







\section{Introduction}\label{introduction}
The problem of quantum mechanical degeneracy in the 2:1 anisotropic oscillator with commensurable frequencies has been studied in great mathematical detail \cite{PhysRev.57.641, demkov, doi:10.1063/1.1665226, doi:10.1063/1.1666379, doi:10.1063/1.523168}, and more recently a complete algebraic description of the symmetry generators of the quantum anisotropic oscillator with commensurable frequencies has been completed \cite{BONATSOS1999537}. The 2:1 oscillator in particular has been studied in other branches of mathematical physics. It is, for example, the only case of the two-dimensional quantum simple harmonic oscillator (other than the isotropic case) where the system is second order super integrable \cite{FRIS1965354}. It is separable in both Cartesian and parabolic coordinates and its eigenfunctions in parabolic coordinates are related to the confluent Heun equation \cite{AIHPA_1969__10_3_259_0}. Dunkl operator generalisations of the problem have also been considered \cite{vinet}. 

In experimental settings, the understanding of multidimensional quantum anisotropic oscillators has been important in describing the states of deformed nuclei \cite{PhysRevLett.64.2623, PhysRevLett.65.301}, and the two-dimensional model has been suggested to be of use in semiconductor physics \cite{asdd} by modelling electrons moving on an anisotropic lattice \cite{Qiong_Gui_2002}. In mesoscopic physics it has been experimentally verified that mode locking of three-dimensional coherent waves of the anisotropic oscillator on parametric Lissajous surfaces form a nearly complete devil's staircase \cite{PhysRevLett.96.213902}.

Multidimensional coherent states have also attracted some interest in the development of their general formalism \cite{Novaes_20022} and in studying their classical limits \cite{Li_2018, Kumar_2008}. Work on coherent states for the anisotropic oscillator is limited, though some have been written down by ansatz \cite{Chen_2003}, they are not presented with ladder operators or a resolution of the identity. Coherent states have many desirable properties in both the mathematical and physical sense. A particular curiosity is their closeness to their classical counterparts be it through minimised uncertainty relations (a canonical coherent state limit) or as we shall find in the example of the anisotropic oscillator, this closeness may be quantified by the reproduction of Lissajous figures in the probability distribution of purely quantum states.

In this paper we develop a scheme for constructing $\mathfrak{su}(2)$-like coherent states for the 2:1 two-dimensional anisotropic harmonic oscillator. The principle idea is to extend the construction from the isotropic setting presented in \cite{isoand}, where ladder operators were defined to organise the degenerate spectrum into a non-degenerate increasing spectrum, to the more general case where the frequencies between the two modes are different, but still commensurable. In doing so we need to introduce non-linear modifications to the ladder operators intended to remove the degeneracy, these operators also mix the two modes. We define a particular natural choice for these ladder operators which properly select all contributing states to the organised spectrum of the 2:1 oscillator.

Degeneracy in quantum systems must be addressed in order to properly define their corresponding coherent states \cite{PhysRevA.64.042104}. This is because generalised definitions of coherent states rely on having a properly ordered spectrum. Suppose we have a non-degenerate discrete spectrum ordered in the following way
\begin{equation}
E_0<E_1<E_2<\ldots<E_k,
\end{equation}
where $k$ may be finite or infinite, and their associated eigenstates $\{\ket{n}\}$ form an orthonormal basis
\begin{equation}
\sum_{n=0}^k \ket{n}\bra{n}=\mathds{1}.
\end{equation}
Klauder showed that generalized coherent states may be expressed as \cite{Klauder_1996}
\begin{equation}\label{rho}
\ket{\xi, \gamma}=M(\xi^2)\sum_{n=0}^k\frac{\xi^n}{\sqrt{\rho_n}}e^{-i \gamma E_n}\ket{n},
\end{equation}
where $0\leq \xi <\infty$ and $-\infty<\gamma <\infty$. The normalisation $M(\xi^2)$ is chosen such that $\bra{\xi, \gamma}\ket{\xi, \gamma}=1$, and the parameters $\rho_n$ are solutions to a moment equation with a positive weight function $k(u)$,
\begin{equation}
\rho_n=\int_0^\infty \mathrm{d}u\, u^n M^2(u)k(u).
\end{equation}
Furthermore we have the completeness relation
\begin{equation}\label{resofid}
\int \mathrm{d} \mu (\xi,\gamma) \ket{\xi, \gamma}\bra{\xi, \gamma} = \mathds{1},
\end{equation}
where integration on the measure $\mathrm{d} \mu (\xi,\gamma)$ is defined as
\begin{equation}\label{measure}
\int \mathrm{d} \mu (\xi,\gamma)=\lim_{\Gamma \rightarrow \infty} \frac{1}{2 \Gamma} \int_0^\infty \mathrm{d}\xi^2\, k(\xi^2)\int_{-\Gamma}^\Gamma \mathrm{d}\gamma.
\end{equation}

The key property here is the resolution of the identity in equation \eqref{resofid}. A resolution of the identity means we have a complete family of continuously parametrised states and therefore may represent any other state in the system in terms of the family of coherent states. This is often considered a defining property of generalised coherent states and it fundamentally holds under this construction due to the organisation of the spectrum.

If on the other hand the spectrum is degenerate we lose the property of the resolution of the identity because the phase factor  $e^{-i \gamma E_n}$ is degenerate. Performing the phase integrations integrations in \eqref{measure} will lead to terms of the form
\begin{equation}
\lim_{\Gamma \rightarrow \infty}\frac{1}{2\Gamma}\int_{-\Gamma}^\Gamma \mathrm{d}\gamma\, e^{i\gamma(E_n-E_m)}
\end{equation}
which should yield a term proportional to $\delta_{nm}$ for non-degenerate energies, but in the event of degeneracy there exists $n\not=m$ such that $E_n=E_m$ and thus we lose proper interpretation of the integral. Work on addressing coherent states for degenerate spectra under this formalism has been completed by Crawford \cite{PhysRevA.62.012104} where a factor of the degree of degeneracy is included in \eqref{rho}, as well as Fox and Choi \cite{PhysRevA.64.042104} in which they add complex phases to the degenerate contributions in order to recover a well defined identity operator. More recently sets of ladder operators were defined in the example of the two-dimensional particle in a box \cite{doi:10.1063/1.2435596} which describes a framework whereby the ladder operators act on the basis states and properly account for the degeneracy.

An ansatz for the form of the coherent states of the anisotropic quantum oscillator has been made \cite{Chen_2003}, for the example of the 2:1 oscillator they would read
\begin{equation}\label{crep}
\ket{\varphi_\nu}=\sum_{k=0}^\nu\alpha^k \beta^{\nu-k}\sqrt{\binom{\nu}{k}}\ket{k,2(\nu-k)},
\end{equation}
which generalises the form of the $\mathfrak{su}(2)$ coherent states of the isotropic oscillator. The states \eqref{crep} produce Lissajous figures in their spatial distribution. It is clear however that these states cannot resolve the identity on the full Hilbert space of states because they miss out states with an odd number in the second mode, $\ket{k,2(\nu-k)+1}$, in their construction. And while it is possible to retroactively fit ladder operators to generate these states, they often include terms such as inverse square roots of the number operator and as such are only well defined on particular representations.

This paper will be organised as follows. In Section \ref{symm} we will introduce the 2:1 two-dimensional anisotropic harmonic oscillator and describe the set up of the problem including the degeneracy in the spectrum and how we wish to organise it to remove the degeneracy. We define sets of states on the non-degenerate spectrum as linear superpositions of states with equal total energy. In section \ref{sec3} we introduce some general definitions that we require the generalised ladder operators to satisfy. We explicitly define a set of non-linear ladder operators $\mathcal{A}^+,\mathcal{A}^-$, compute all zero modes associated with the operator $\mathcal{A}^-$ and define chains of states associated to each zero mode by the action of $\mathcal{A}^+$. Section \ref{secprin} is devoted to the principle chain of states generated from the lowest energy zero mode, which corresponds to the ground state $\ket{0,0}$. We show that they modify the form of the binomial coefficient of the $\mathfrak{su}(2)$ coherent states, and we find that they admit a resolution of the identity. Following this in section \ref{uncertainty} we discuss the uncertainty relations for the principle chain of states and find that the uncertainties in each mode are intrinsically connected due to the parameters introduced in their construction. Finally, we conclude in section \ref{conc} by discussing the possibility of generalising the techniques developed in this paper to other systems.

\section{Bosonic states and degenerate spectra}\label{symm}
The quantum hamiltonian for the 2:1 oscillator whose frequency in its $a$ mode is twice that of its frequency in its $b$ mode is given in terms of the canonical position and momentum operators $\hat{Q}_i, \hat{P}_i$ respectively by
\begin{equation}\label{mainform}
 \hat{H}=\frac{1}{2}\hat{P}_a^2+\frac{1}{2}\hat{P}_b^2+2\hat{Q}_a^2 +\frac{1}{2}\hat{Q}_b^2,
\end{equation}
where we have set the dimensionful quantities $\hbar,\omega,m=1$, Planck's constant, the common frequency and the geometric mean mass respectively. The position and momentum operators satisfy the canonical commutation relations
\begin{equation}
[Q_a,P_b]=i \delta_{ab}\mathds{1},\quad [Q_a,Q_b]=0, \quad [P_a,P_b]=0,
\end{equation}
where $\mathds{1}$ is the identity operator. For our purposes we are interested in the corresponding ladder operator formalism, achieved by defining the operators
\begin{equation}\label{ladders}
\begin{split}
 &a^- =\left(\hat{Q}_a + \frac{i}{2} \hat{P}_a\right), \\
 &b^- =\frac{1}{\sqrt{2}}\left(\hat{Q}_b + i \hat{P}_b\right),
\end{split}
\qquad
\begin{split}
 &a^+ =\left(\hat{Q}_a - \frac{i}{2} \hat{P}_a\right),\\
 &b^+ =\frac{1}{\sqrt{2}}\left(\hat{Q}_b - i \hat{P}_b\right),
\end{split} 
\end{equation}
which can be used to rewrite the hamiltonian \eqref{mainform} as
\begin{equation}\label{ladform}
\hat{H}=2a^+a^- + b^+b^- +\frac{3}{2}\mathds{1}.
\end{equation}
The operators \eqref{ladders} and \eqref{ladform} satisfy the canonical commutation relations in the $a$ and $b$ modes separately,
\begin{equation}
\begin{split}
 &[a^-,a^+]=\mathds{1}, \\
 &[\hat{H},a^{\pm}]=\pm 2 a^\pm,
\end{split}
\qquad
\begin{split}
 &[b^-,b^+]=\mathds{1},\\
 & [\hat{H},b^{\pm}]=\pm  b^\pm,
\end{split} 
\end{equation}
with other commutators vanishing. Notice that in our definitions we have absorbed the difference in frequencies into the definition of the ladder operators and not into their commutation relations, so we preserve $[a^-,a^+]=[b^-,b^+]=\mathds{1}$.

The energy eigenstates are solutions to the time-independent Schr\"odinger equation
\begin{equation}
\hat{H}\ket{n,m}=E_{n,m}\ket{n,m},
\end{equation}
where $n,m \in \mathds{Z}^{\geq0}$. The ladder operators \eqref{ladders} have the following actions on the energy eigenstates
\begin{equation}
\begin{split}
 &a^-\ket{n,m}=\sqrt{n}\ket{n-1,m},\\
 &b^-\ket{n,m}=\sqrt{m}\ket{n,m-1},
\end{split}
\qquad
\begin{split}
 &a^+\ket{n,m}=\sqrt{n+1}\ket{n+1,m},\\
 &b^+\ket{n,m}=\sqrt{m+1}\ket{n,m+1},
\end{split} 
\end{equation}
and the eigenvalues $E_{n,m}$ are given by
\begin{equation}\label{degen}
E_{n,m}=2n+m+\frac{3}{2}.
\end{equation}
Equation \eqref{degen} is our first descriptor of the degeneracy present in the 2:1 oscillator, it is equivalent to a simple problem in number theory, namely, for $\nu, n, m\in \mathds{Z}^{\geq0}$, what values can $n$ and $m$ take satisfying the following equation
\begin{equation}\label{number}
\nu=2n+m.
\end{equation}
In terms of $\nu$ the spectrum $E_{n,m}=E_\nu = \nu+\frac{3}{2}$ is well organised
\begin{equation}
E_0<E_1<E_2<\ldots,
\end{equation}
and the states with energy $E_\nu$ will be expressed as linear superpositions of all states with energy $E_{n,m}=E_\nu$ as
\begin{equation}\label{alllam}
\ket{\varphi_\nu}=\sum_{k=0}^{\floor{\frac{\nu}{2}}} \Lambda_k(\nu, \{\alpha_i\}) \ket{k,\nu-2k}.
\end{equation}
\begin{table}[H]%
\begin{center}%
\begin{tabular}{ c|c }%
  $\ket{\bm{\varphi_\nu}}$& $\textrm{\textbf{contributing states}}$\\%
 \cline{1-2}%
  $\ket{\varphi_0}$ & $\ket{0,0}$\\%
  $\ket{\varphi_1}$ & $\ket{0,1}$ \\%
  $\ket{\varphi_2}$ & $\ket{1,0}, \ket{0,2}$\\%
  $\ket{\varphi_3}$ & $\ket{1,1}, \ket{0,3}$\\%
  $\vdots$ & $\vdots$ \\%
  $\ket{\varphi_\nu}$ & $\left\{\ket{k,\nu-2k}\mid k = 0,1,2,\ldots,\floor{\frac{\nu}{2}}\right\} $ %
\end{tabular}%
\end{center}\caption{States $\ket{n,m}$ contributing to a state $\ket{\varphi_\nu}$ with energy $E_\nu$.}%
\label{table1}
\end{table}%

Here $\Lambda_k(\nu, \{\alpha_i\})$ are the complex expansion coefficients, with complex conjugate $\bar{\Lambda}_{k}(\nu, \{\alpha_i\})$, and they may depend on additional variables $\{\alpha_i\}, i \in \{1,2,\ldots, m\}$. The coefficients represent the weights we attribute to each Fock basis state, after which we sum over the number $k$ to produce an averaged contribution to a state $\ket{\varphi_\nu}$.

We require that the states \eqref{alllam} satisfy the following properties: normalisation
\begin{equation}
\sum_{k=0}^{\floor{\frac{\nu}{2}}} \bar{\Lambda}_k(\nu, \{\alpha_i\}) \Lambda_k(\nu, \{\alpha_i\}) =1,
\end{equation}
and the completeness relation
\begin{equation}\label{comp}
\int_{\mathds{C}^m}\mathrm{d}\mu(\{\alpha_i\})\, \bar{\Lambda}_{k'}(\nu, \{\alpha_i\}) \Lambda_k(\nu, \{\alpha_i\})=\delta_{k' k},
\end{equation}
for some measure $\mathrm{d}\mu(\{\alpha_i\})$ such that they resolve the identity on the subspace indexed by $\nu$
\begin{equation}\label{ress}
\int_{\mathds{C}^m}\mathrm{d}\mu(\{\alpha_i\})\ket{\varphi_\nu}\bra{\varphi_\nu}=\mathds{1}_{\mathcal{\nu}},
\end{equation}
and over the total Hilbert space via
\begin{equation}
\sum_{\nu=0}^\infty \mathds{1}_{\mathcal{\nu}}=\mathds{1}_{\mathcal{H}}.
\end{equation}
Here $\mathds{1}_{\mathcal{H}}=\sum_{n,m}^\infty \ket{n,m}\bra{n,m}$ refers to the identity operator on the full Hilbert space of states. This choice of partitioning the space of states $\{\ket{n,m}\}$ is the most natural when considering the 2:1 oscillator.

There is some freedom in defining the set of states $\left\{\ket{\varphi_\nu}\right\}$ as the contributing states can be weighted in any way such that \eqref{comp} and \eqref{ress} are satisfied, these represent a proper probabilistic interpretation and completeness, respectively. By defining sets of ladder operators which correctly pick out the states in table \ref{table1}, the set of coefficients defining the states $\left\{\ket{\varphi_\nu}\right\}$ are predetermined. These can always be normalised, the pertinent calculation is to check that a measure can be found for \eqref{ress}. 

\section{Ladder operators and chains of states}\label{sec3}
Now we turn our attention to the ladder operators which will generate the states \eqref{alllam}. A defining feature of the generalised ladder operators of our system will be
\begin{equation}\label{hamcom}
 [\hat{H},\mathcal{A}^+]=\mathcal{A}^+, \quad [\hat{H},\mathcal{A}^-]=-\mathcal{A}^-,
\end{equation}
this ensures that if $\ket{\psi_\nu}$ is an eigenstate of $\hat{H}$ then so are $\mathcal{A}^-\ket{\psi_\nu}$ and $\mathcal{A}^+\ket{\psi_\nu}$. Operators in more than one variable allow for infinitely many zero modes, to this end, the generalised annihilation operator, $\mathcal{A}^-$, admits the following zero modes 
\begin{equation}\label{zero}
 \mathcal{A}^-\ket{\varphi_0^{(n)}}=0,
\end{equation}
where the index $(n)$ enumerates zero modes and we define the principle zero mode $\ket{\varphi_0^{(0)}}=\ket{0,0}$ to be the ground state of \eqref{ladform}. The Fock spaces (chains of states) associated to each zero mode $\ket{\varphi_0^{(n)}}$ are generated by
\begin{equation}\label{creation}
 \ket{\varphi_\nu^{(n)}}=\frac{1}{\sqrt{[f(\nu)]!}}(\mathcal{A}^+)^\nu \ket{\varphi_0^{(n)}},
\end{equation}
where $\nu$ describes the $\nu$-th state in the chain and the states generated from the principle zero mode, $\ket{\varphi_\nu^{(0)}}$, define the principle chain of states. The function $f(\nu)$ is chosen such that the states are normalised $\bra{\varphi_\nu^{(n)}} \ket{\varphi_\nu^{(n)}}=1$, and the action on an intermediate state is given by
\begin{equation}\label{condit}
\mathcal{A}^+ \ket{\varphi_\nu^{(n)}} =\sqrt{f(\nu+1)}\ket{\varphi_{\nu+1}^{(n)}}.
\end{equation}
Because we are interested in the constructive generation of states we do not define the action of $\mathcal{A}^-$ on an intermediate state because we will choose $\mathcal{A}^-=(\mathcal{A}^+)^\dagger$ and in general its action will lead to a superposition of states from different chains. Much of the framework described follows from the defining features of the ladder operators for the one-dimensional harmonic oscillator, but we relax the canonical commutation relation such that
\begin{equation}
[\mathcal{A}^-,\mathcal{A}^+] \not = \mathds{1},
\end{equation}
the commutation relation will not be canonical, this allows us to define generalised ladder operators as non-linear combinations in $a^-, a^+, b^-$, and $b^+$.

With the framework described we find that a suitable pair of ladder operators are
\begin{equation}\label{operators}
 \mathcal{A}^+ = \alpha b^+ + \beta a^+ b^-, \quad \mathcal{A}^- = \bar{\alpha} b^- + \bar{\beta} a^- b^+.
\end{equation}
Their commutator yields
\begin{equation}
[\mathcal{A}^-,\mathcal{A}^+]=\abs{\alpha}^2\mathds{1}+\abs{\beta}^2(b^+ b^- - a^+ a^-),
\end{equation}
and it can be verified that these operators obey \eqref{hamcom} with hamiltonian \eqref{ladform}. Firstly we categorise all of the zero modes of the operator $\mathcal{A}^-$ by solving \eqref{zero} in the basis \eqref{alllam}
\begin{equation}
\ket{\psi_0^{(n)}}=\sum_{j=0}^{\floor{\frac{n}{2}}} \gamma_j^{(n)}(\alpha,\beta)\ket{j,n-2j}, \quad n\in \mathds{Z}^{\geq 0},
\end{equation}
assuming $ \gamma_j^{(n)}(\alpha,\beta) \not = 0$. Considering $n\rightarrow 2n'$ even, equation \eqref{zero} is explicitly
\begin{equation}\label{zerop}
\begin{split}
\sum_{j=0}^{n'} \Big(& \gamma_j^{(2n')}(\alpha,\beta) \bar{\alpha}\sqrt{2(n'-j)} \ket{j,2n'-2j-1}\\
&+ \gamma_j^{(2n')}(\alpha,\beta) \bar{\beta}\sqrt{j-1}\sqrt{2(n'-j)+1}\ket{j-1,2n'-2j+1} \Big)=0.
\end{split}
\end{equation}
After relabelling the second summation index $j\rightarrow j+1$ in \eqref{zerop} we find that the coefficients $ \gamma_j^{(2n')}(\alpha,\beta)$ satisfy the following recursion relation on the index $j$,
\begin{equation}
\gamma_{j+1}^{(2n')}(\alpha,\beta)=-\frac{\bar{\alpha}}{\bar{\beta}}\frac{\sqrt{2(n'-j)}}{\sqrt{j+1}\sqrt{2(n'-j)-1}}\gamma_j^{(2n')}(\alpha,\beta).
\end{equation}
This can be solved straightforwardly with the definition $\gamma_0^{(2n')}(\alpha,\beta)=1$ (this is just an overall multiplicative constant) to give
\begin{equation}\label{even}
\begin{split}
\gamma_{j}^{(2n')}(\alpha,\beta)&=\left(-\frac{\bar{\alpha}}{\bar{\beta}}\right)^{j}\prod_{k=0}^{j-1} \frac{\sqrt{2(n'-j+k+1)}}{\sqrt{j-k}\sqrt{2(n'-j+k)+1}},\\
&=\left(-2\frac{\bar{\alpha}}{\bar{\beta}}\right)^{j}\left(\frac{n'!}{(n'-j)!} \right)\sqrt{\frac{(2(n'-j))!}{j!(2n')!}}, \quad j \in \{0, 1, \ldots, n' \}.
\end{split}
\end{equation}

Repeating the procedure for  $n\rightarrow2n'+1$ odd,  
\begin{equation}
\ket{\psi_0^{(2n'+1)}}=\sum_{j=0}^{n'} \gamma_j^{(2n'+1)}(\alpha,\beta)\ket{j,2(n'-j)+1},
\end{equation}
solving equation \eqref{zero} we find a similar recursion relation to the case where $n$ is even, but with an added constraint,
\begin{equation}
\begin{split}
 \gamma_j^{(2n'+1)}(\alpha,\beta) \bar{\alpha}\sqrt{2(n'-j)+1}&+ \gamma_{j+1}^{(2n'+1)}(\alpha,\beta)\bar{\beta}\sqrt{j+1}\sqrt{2(n'-j)}=0,\\
 &\gamma_n^{(2n'+1)}(\alpha,\beta)=0.
 \end{split}
\end{equation}
The last constraint $\gamma_n^{(2n'+1)}(\alpha,\beta)=0$ implies that all terms in the recursion vanish and as a result we have no zero modes associated with odd values of $n$, that is there are no zero modes of $\mathcal{A}^-$ associated to odd values of $\nu$. Thus, all normalised zero modes are given by
\begin{equation}\label{normgs}
 \ket{\varphi_0^{(2n)}}=\frac{1}{\sqrt{\mathcal{N}^{(2n)}_0(\alpha,\beta)}}\sum_{j=0}^{n} \gamma_j^{(2n)}(\alpha,\beta)\ket{j,2(n-j)},\quad  n \in \mathds{Z}^{\geq 0},
\end{equation}
with coefficients
\begin{equation}\label{gamcoef}
\gamma_j^{(2n)}(\alpha,\beta)=
 \left(-2\frac{\bar{\alpha}}{\bar{\beta}}\right)^{j}\left(\frac{n!}{(n-j)!} \right)\sqrt{\frac{(2(n-j))!}{j!(2n)!} }, \quad j \in \left\{0,1, \ldots, n \right\}.
\end{equation}
We have introduced the normalisation function $\mathcal{N}^{(2n)}_0(\alpha,\beta)$ which ensures that $\bra{\varphi_0^{(2n)}}\ket{\varphi_0^{(2n)}}=1$, $\forall n\in \mathds{Z}^{\geq 0}$. It is given in terms of the coefficients $\gamma_j^{(2n)}(\alpha,\beta)$ by
\begin{equation}
\mathcal{N}^{(2n)}_0(\alpha,\beta)=\sum_{j=0}^{n}\abs{\gamma_j^{(2n)}(\alpha,\beta)}^2.
\end{equation}

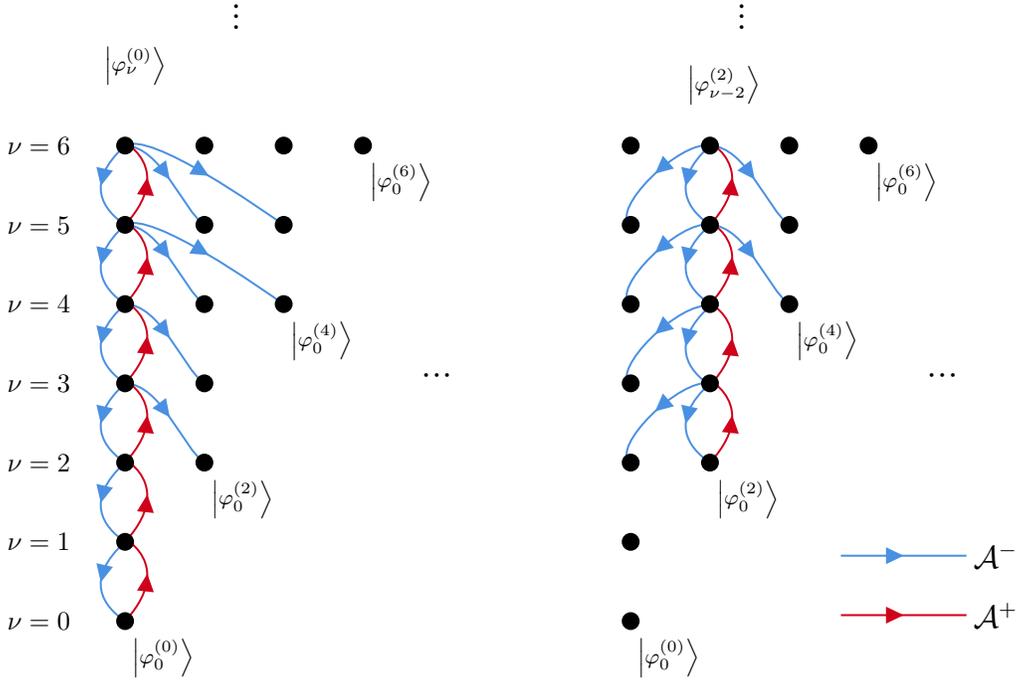
\begin{figure}[H]

\tikzset{every picture/.style={line width=0.75pt}} 

\begin{tikzpicture}[x=0.75pt,y=0.75pt,yscale=-1,xscale=1]

\draw [color={rgb, 255:red, 74; green, 144; blue, 226 }  ,draw opacity=1 ][line width=0.75]    (59,334) .. controls (38,353.17) and (48,366.5) .. (59,374) ;
\draw [shift={(46.75,354.42)}, rotate = 282.72] [fill={rgb, 255:red, 74; green, 144; blue, 226 }  ,fill opacity=1 ][line width=0.08]  [draw opacity=0] (8.93,-4.29) -- (0,0) -- (8.93,4.29) -- cycle    ;
\draw [color={rgb, 255:red, 74; green, 144; blue, 226 }  ,draw opacity=1 ][line width=0.75]    (59,294) .. controls (38,313.17) and (48,326.5) .. (59,334) ;
\draw [shift={(46.75,314.42)}, rotate = 282.72] [fill={rgb, 255:red, 74; green, 144; blue, 226 }  ,fill opacity=1 ][line width=0.08]  [draw opacity=0] (8.93,-4.29) -- (0,0) -- (8.93,4.29) -- cycle    ;
\draw [color={rgb, 255:red, 74; green, 144; blue, 226 }  ,draw opacity=1 ][line width=0.75]    (59,294) .. controls (77.33,294.5) and (88,326.5) .. (99,334) ;
\draw [shift={(81.56,311.11)}, rotate = 233.12] [fill={rgb, 255:red, 74; green, 144; blue, 226 }  ,fill opacity=1 ][line width=0.08]  [draw opacity=0] (8.93,-4.29) -- (0,0) -- (8.93,4.29) -- cycle    ;
\draw [color={rgb, 255:red, 74; green, 144; blue, 226 }  ,draw opacity=1 ][line width=0.75]    (59,254) .. controls (38,273.17) and (48,286.5) .. (59,294) ;
\draw [shift={(46.75,274.42)}, rotate = 282.72] [fill={rgb, 255:red, 74; green, 144; blue, 226 }  ,fill opacity=1 ][line width=0.08]  [draw opacity=0] (8.93,-4.29) -- (0,0) -- (8.93,4.29) -- cycle    ;
\draw [color={rgb, 255:red, 74; green, 144; blue, 226 }  ,draw opacity=1 ][line width=0.75]    (59,254) .. controls (77.33,254.5) and (88,286.5) .. (99,294) ;
\draw [shift={(81.56,271.11)}, rotate = 233.12] [fill={rgb, 255:red, 74; green, 144; blue, 226 }  ,fill opacity=1 ][line width=0.08]  [draw opacity=0] (8.93,-4.29) -- (0,0) -- (8.93,4.29) -- cycle    ;
\draw [color={rgb, 255:red, 74; green, 144; blue, 226 }  ,draw opacity=1 ][line width=0.75]    (59,214) .. controls (77.33,214.5) and (88,246.5) .. (99,254) ;
\draw [shift={(81.56,231.11)}, rotate = 233.12] [fill={rgb, 255:red, 74; green, 144; blue, 226 }  ,fill opacity=1 ][line width=0.08]  [draw opacity=0] (8.93,-4.29) -- (0,0) -- (8.93,4.29) -- cycle    ;
\draw [color={rgb, 255:red, 74; green, 144; blue, 226 }  ,draw opacity=1 ][line width=0.75]    (59,214) .. controls (38,233.17) and (48,246.5) .. (59,254) ;
\draw [shift={(46.75,234.42)}, rotate = 282.72] [fill={rgb, 255:red, 74; green, 144; blue, 226 }  ,fill opacity=1 ][line width=0.08]  [draw opacity=0] (8.93,-4.29) -- (0,0) -- (8.93,4.29) -- cycle    ;
\draw [color={rgb, 255:red, 74; green, 144; blue, 226 }  ,draw opacity=1 ][line width=0.75]    (59,214) .. controls (72,206.5) and (128,246.5) .. (139,254) ;
\draw [shift={(101.28,229.28)}, rotate = 210.25] [fill={rgb, 255:red, 74; green, 144; blue, 226 }  ,fill opacity=1 ][line width=0.08]  [draw opacity=0] (8.93,-4.29) -- (0,0) -- (8.93,4.29) -- cycle    ;
\draw [color={rgb, 255:red, 74; green, 144; blue, 226 }  ,draw opacity=1 ][line width=0.75]    (59,174) .. controls (72,166.5) and (128,206.5) .. (139,214) ;
\draw [shift={(101.28,189.28)}, rotate = 210.25] [fill={rgb, 255:red, 74; green, 144; blue, 226 }  ,fill opacity=1 ][line width=0.08]  [draw opacity=0] (8.93,-4.29) -- (0,0) -- (8.93,4.29) -- cycle    ;
\draw [color={rgb, 255:red, 74; green, 144; blue, 226 }  ,draw opacity=1 ][line width=0.75]    (59,174) .. controls (77.33,174.5) and (88,206.5) .. (99,214) ;
\draw [shift={(81.56,191.11)}, rotate = 233.12] [fill={rgb, 255:red, 74; green, 144; blue, 226 }  ,fill opacity=1 ][line width=0.08]  [draw opacity=0] (8.93,-4.29) -- (0,0) -- (8.93,4.29) -- cycle    ;
\draw [color={rgb, 255:red, 74; green, 144; blue, 226 }  ,draw opacity=1 ][line width=0.75]    (59,174) .. controls (38,193.17) and (48,206.5) .. (59,214) ;
\draw [shift={(46.75,194.42)}, rotate = 282.72] [fill={rgb, 255:red, 74; green, 144; blue, 226 }  ,fill opacity=1 ][line width=0.08]  [draw opacity=0] (8.93,-4.29) -- (0,0) -- (8.93,4.29) -- cycle    ;
\draw  [fill={rgb, 255:red, 0; green, 0; blue, 0 }  ,fill opacity=1 ] (95,334) .. controls (95,331.79) and (96.79,330) .. (99,330) .. controls (101.21,330) and (103,331.79) .. (103,334) .. controls (103,336.21) and (101.21,338) .. (99,338) .. controls (96.79,338) and (95,336.21) .. (95,334) -- cycle ;
\draw  [fill={rgb, 255:red, 0; green, 0; blue, 0 }  ,fill opacity=1 ] (95,294) .. controls (95,291.79) and (96.79,290) .. (99,290) .. controls (101.21,290) and (103,291.79) .. (103,294) .. controls (103,296.21) and (101.21,298) .. (99,298) .. controls (96.79,298) and (95,296.21) .. (95,294) -- cycle ;
\draw  [fill={rgb, 255:red, 0; green, 0; blue, 0 }  ,fill opacity=1 ] (95,254) .. controls (95,251.79) and (96.79,250) .. (99,250) .. controls (101.21,250) and (103,251.79) .. (103,254) .. controls (103,256.21) and (101.21,258) .. (99,258) .. controls (96.79,258) and (95,256.21) .. (95,254) -- cycle ;
\draw  [fill={rgb, 255:red, 0; green, 0; blue, 0 }  ,fill opacity=1 ] (95,214) .. controls (95,211.79) and (96.79,210) .. (99,210) .. controls (101.21,210) and (103,211.79) .. (103,214) .. controls (103,216.21) and (101.21,218) .. (99,218) .. controls (96.79,218) and (95,216.21) .. (95,214) -- cycle ;

\draw  [fill={rgb, 255:red, 0; green, 0; blue, 0 }  ,fill opacity=1 ] (95,174) .. controls (95,171.79) and (96.79,170) .. (99,170) .. controls (101.21,170) and (103,171.79) .. (103,174) .. controls (103,176.21) and (101.21,178) .. (99,178) .. controls (96.79,178) and (95,176.21) .. (95,174) -- cycle ;
\draw  [fill={rgb, 255:red, 0; green, 0; blue, 0 }  ,fill opacity=1 ] (135,254) .. controls (135,251.79) and (136.79,250) .. (139,250) .. controls (141.21,250) and (143,251.79) .. (143,254) .. controls (143,256.21) and (141.21,258) .. (139,258) .. controls (136.79,258) and (135,256.21) .. (135,254) -- cycle ;
\draw  [fill={rgb, 255:red, 0; green, 0; blue, 0 }  ,fill opacity=1 ] (135,214) .. controls (135,211.79) and (136.79,210) .. (139,210) .. controls (141.21,210) and (143,211.79) .. (143,214) .. controls (143,216.21) and (141.21,218) .. (139,218) .. controls (136.79,218) and (135,216.21) .. (135,214) -- cycle ;
\draw  [fill={rgb, 255:red, 0; green, 0; blue, 0 }  ,fill opacity=1 ] (135,174) .. controls (135,171.79) and (136.79,170) .. (139,170) .. controls (141.21,170) and (143,171.79) .. (143,174) .. controls (143,176.21) and (141.21,178) .. (139,178) .. controls (136.79,178) and (135,176.21) .. (135,174) -- cycle ;
\draw  [fill={rgb, 255:red, 0; green, 0; blue, 0 }  ,fill opacity=1 ] (175,174) .. controls (175,171.79) and (176.79,170) .. (179,170) .. controls (181.21,170) and (183,171.79) .. (183,174) .. controls (183,176.21) and (181.21,178) .. (179,178) .. controls (176.79,178) and (175,176.21) .. (175,174) -- cycle ;
\draw [color={rgb, 255:red, 208; green, 2; blue, 27 }  ,draw opacity=1 ][line width=0.75]    (59,414) .. controls (89,381.17) and (48,366.5) .. (59,374) ;
\draw [shift={(70.19,390.12)}, rotate = 455.96] [fill={rgb, 255:red, 208; green, 2; blue, 27 }  ,fill opacity=1 ][line width=0.08]  [draw opacity=0] (8.93,-4.29) -- (0,0) -- (8.93,4.29) -- cycle    ;
\draw [color={rgb, 255:red, 208; green, 2; blue, 27 }  ,draw opacity=1 ][line width=0.75]    (59,374) .. controls (89,341.17) and (48,326.5) .. (59,334) ;
\draw [shift={(70.19,350.12)}, rotate = 455.96] [fill={rgb, 255:red, 208; green, 2; blue, 27 }  ,fill opacity=1 ][line width=0.08]  [draw opacity=0] (8.93,-4.29) -- (0,0) -- (8.93,4.29) -- cycle    ;
\draw [color={rgb, 255:red, 208; green, 2; blue, 27 }  ,draw opacity=1 ][line width=0.75]    (59,334) .. controls (89,301.17) and (48,286.5) .. (59,294) ;
\draw [shift={(70.19,310.12)}, rotate = 455.96] [fill={rgb, 255:red, 208; green, 2; blue, 27 }  ,fill opacity=1 ][line width=0.08]  [draw opacity=0] (8.93,-4.29) -- (0,0) -- (8.93,4.29) -- cycle    ;
\draw [color={rgb, 255:red, 208; green, 2; blue, 27 }  ,draw opacity=1 ][line width=0.75]    (59,294) .. controls (89,261.17) and (48,246.5) .. (59,254) ;
\draw [shift={(70.19,270.12)}, rotate = 455.96] [fill={rgb, 255:red, 208; green, 2; blue, 27 }  ,fill opacity=1 ][line width=0.08]  [draw opacity=0] (8.93,-4.29) -- (0,0) -- (8.93,4.29) -- cycle    ;
\draw [color={rgb, 255:red, 208; green, 2; blue, 27 }  ,draw opacity=1 ][line width=0.75]    (59,254) .. controls (89,221.17) and (48,206.5) .. (59,214) ;
\draw [shift={(70.19,230.12)}, rotate = 455.96] [fill={rgb, 255:red, 208; green, 2; blue, 27 }  ,fill opacity=1 ][line width=0.08]  [draw opacity=0] (8.93,-4.29) -- (0,0) -- (8.93,4.29) -- cycle    ;
\draw [color={rgb, 255:red, 208; green, 2; blue, 27 }  ,draw opacity=1 ][line width=0.75]    (59,214) .. controls (89,181.17) and (48,166.5) .. (59,174) ;
\draw [shift={(70.19,190.12)}, rotate = 455.96] [fill={rgb, 255:red, 208; green, 2; blue, 27 }  ,fill opacity=1 ][line width=0.08]  [draw opacity=0] (8.93,-4.29) -- (0,0) -- (8.93,4.29) -- cycle    ;

\draw  [fill={rgb, 255:red, 0; green, 0; blue, 0 }  ,fill opacity=1 ] (55,294) .. controls (55,291.79) and (56.79,290) .. (59,290) .. controls (61.21,290) and (63,291.79) .. (63,294) .. controls (63,296.21) and (61.21,298) .. (59,298) .. controls (56.79,298) and (55,296.21) .. (55,294) -- cycle ;
\draw  [fill={rgb, 255:red, 0; green, 0; blue, 0 }  ,fill opacity=1 ] (55,254) .. controls (55,251.79) and (56.79,250) .. (59,250) .. controls (61.21,250) and (63,251.79) .. (63,254) .. controls (63,256.21) and (61.21,258) .. (59,258) .. controls (56.79,258) and (55,256.21) .. (55,254) -- cycle ;
\draw  [fill={rgb, 255:red, 0; green, 0; blue, 0 }  ,fill opacity=1 ] (55,214) .. controls (55,211.79) and (56.79,210) .. (59,210) .. controls (61.21,210) and (63,211.79) .. (63,214) .. controls (63,216.21) and (61.21,218) .. (59,218) .. controls (56.79,218) and (55,216.21) .. (55,214) -- cycle ;

\draw  [fill={rgb, 255:red, 0; green, 0; blue, 0 }  ,fill opacity=1 ] (55,334) .. controls (55,331.79) and (56.79,330) .. (59,330) .. controls (61.21,330) and (63,331.79) .. (63,334) .. controls (63,336.21) and (61.21,338) .. (59,338) .. controls (56.79,338) and (55,336.21) .. (55,334) -- cycle ;
\draw [color={rgb, 255:red, 74; green, 144; blue, 226 }  ,draw opacity=1 ][line width=0.75]    (59,374) .. controls (38,393.17) and (48,406.5) .. (59,414) ;
\draw [shift={(46.75,394.42)}, rotate = 282.72] [fill={rgb, 255:red, 74; green, 144; blue, 226 }  ,fill opacity=1 ][line width=0.08]  [draw opacity=0] (8.93,-4.29) -- (0,0) -- (8.93,4.29) -- cycle    ;
\draw  [fill={rgb, 255:red, 0; green, 0; blue, 0 }  ,fill opacity=1 ] (55,374) .. controls (55,371.79) and (56.79,370) .. (59,370) .. controls (61.21,370) and (63,371.79) .. (63,374) .. controls (63,376.21) and (61.21,378) .. (59,378) .. controls (56.79,378) and (55,376.21) .. (55,374) -- cycle ;
\draw  [fill={rgb, 255:red, 0; green, 0; blue, 0 }  ,fill opacity=1 ] (55,414) .. controls (55,411.79) and (56.79,410) .. (59,410) .. controls (61.21,410) and (63,411.79) .. (63,414) .. controls (63,416.21) and (61.21,418) .. (59,418) .. controls (56.79,418) and (55,416.21) .. (55,414) -- cycle ;
\draw  [fill={rgb, 255:red, 0; green, 0; blue, 0 }  ,fill opacity=1 ] (55,174) .. controls (55,171.79) and (56.79,170) .. (59,170) .. controls (61.21,170) and (63,171.79) .. (63,174) .. controls (63,176.21) and (61.21,178) .. (59,178) .. controls (56.79,178) and (55,176.21) .. (55,174) -- cycle ;
\draw [color={rgb, 255:red, 74; green, 144; blue, 226 }  ,draw opacity=1 ][line width=0.75]    (420,381) -- (483.38,381) ;
\draw [shift={(451.69,381)}, rotate = 180] [fill={rgb, 255:red, 74; green, 144; blue, 226 }  ,fill opacity=1 ][line width=0.08]  [draw opacity=0] (8.93,-4.29) -- (0,0) -- (8.93,4.29) -- cycle    ;
\draw [color={rgb, 255:red, 208; green, 2; blue, 27 }  ,draw opacity=1 ][line width=0.75]    (420,411) -- (483.38,411) ;
\draw [shift={(451.69,411)}, rotate = 180] [fill={rgb, 255:red, 208; green, 2; blue, 27 }  ,fill opacity=1 ][line width=0.08]  [draw opacity=0] (8.93,-4.29) -- (0,0) -- (8.93,4.29) -- cycle    ;

\draw [color={rgb, 255:red, 74; green, 144; blue, 226 }  ,draw opacity=1 ][line width=0.75]    (354,294) .. controls (333,313.17) and (343,326.5) .. (354,334) ;
\draw [shift={(341.75,314.42)}, rotate = 282.72] [fill={rgb, 255:red, 74; green, 144; blue, 226 }  ,fill opacity=1 ][line width=0.08]  [draw opacity=0] (8.93,-4.29) -- (0,0) -- (8.93,4.29) -- cycle    ;
\draw [color={rgb, 255:red, 74; green, 144; blue, 226 }  ,draw opacity=1 ][line width=0.75]    (354,294) .. controls (339,291.17) and (303,326.5) .. (314,334) ;
\draw [shift={(326.29,308.17)}, rotate = 319.18] [fill={rgb, 255:red, 74; green, 144; blue, 226 }  ,fill opacity=1 ][line width=0.08]  [draw opacity=0] (8.93,-4.29) -- (0,0) -- (8.93,4.29) -- cycle    ;
\draw [color={rgb, 255:red, 74; green, 144; blue, 226 }  ,draw opacity=1 ][line width=0.75]    (354,214) .. controls (372.33,214.5) and (383,246.5) .. (394,254) ;
\draw [shift={(376.56,231.11)}, rotate = 233.12] [fill={rgb, 255:red, 74; green, 144; blue, 226 }  ,fill opacity=1 ][line width=0.08]  [draw opacity=0] (8.93,-4.29) -- (0,0) -- (8.93,4.29) -- cycle    ;
\draw [color={rgb, 255:red, 74; green, 144; blue, 226 }  ,draw opacity=1 ][line width=0.75]    (354,174) .. controls (372.33,174.5) and (383,206.5) .. (394,214) ;
\draw [shift={(376.56,191.11)}, rotate = 233.12] [fill={rgb, 255:red, 74; green, 144; blue, 226 }  ,fill opacity=1 ][line width=0.08]  [draw opacity=0] (8.93,-4.29) -- (0,0) -- (8.93,4.29) -- cycle    ;
\draw [color={rgb, 255:red, 74; green, 144; blue, 226 }  ,draw opacity=1 ][line width=0.75]    (354,255) .. controls (333,274.17) and (343,287.5) .. (354,295) ;
\draw [shift={(341.75,275.42)}, rotate = 282.72] [fill={rgb, 255:red, 74; green, 144; blue, 226 }  ,fill opacity=1 ][line width=0.08]  [draw opacity=0] (8.93,-4.29) -- (0,0) -- (8.93,4.29) -- cycle    ;
\draw [color={rgb, 255:red, 74; green, 144; blue, 226 }  ,draw opacity=1 ][line width=0.75]    (354,255) .. controls (339,252.17) and (303,287.5) .. (314,295) ;
\draw [shift={(326.29,269.17)}, rotate = 319.18] [fill={rgb, 255:red, 74; green, 144; blue, 226 }  ,fill opacity=1 ][line width=0.08]  [draw opacity=0] (8.93,-4.29) -- (0,0) -- (8.93,4.29) -- cycle    ;
\draw [color={rgb, 255:red, 74; green, 144; blue, 226 }  ,draw opacity=1 ][line width=0.75]    (354,214) .. controls (333,233.17) and (343,246.5) .. (354,254) ;
\draw [shift={(341.75,234.42)}, rotate = 282.72] [fill={rgb, 255:red, 74; green, 144; blue, 226 }  ,fill opacity=1 ][line width=0.08]  [draw opacity=0] (8.93,-4.29) -- (0,0) -- (8.93,4.29) -- cycle    ;
\draw [color={rgb, 255:red, 74; green, 144; blue, 226 }  ,draw opacity=1 ][line width=0.75]    (354,214) .. controls (339,211.17) and (303,246.5) .. (314,254) ;
\draw [shift={(326.29,228.17)}, rotate = 319.18] [fill={rgb, 255:red, 74; green, 144; blue, 226 }  ,fill opacity=1 ][line width=0.08]  [draw opacity=0] (8.93,-4.29) -- (0,0) -- (8.93,4.29) -- cycle    ;
\draw [color={rgb, 255:red, 74; green, 144; blue, 226 }  ,draw opacity=1 ][line width=0.75]    (354,174) .. controls (333,193.17) and (343,206.5) .. (354,214) ;
\draw [shift={(341.75,194.42)}, rotate = 282.72] [fill={rgb, 255:red, 74; green, 144; blue, 226 }  ,fill opacity=1 ][line width=0.08]  [draw opacity=0] (8.93,-4.29) -- (0,0) -- (8.93,4.29) -- cycle    ;
\draw [color={rgb, 255:red, 74; green, 144; blue, 226 }  ,draw opacity=1 ][line width=0.75]    (354,174) .. controls (339,171.17) and (303,206.5) .. (314,214) ;
\draw [shift={(326.29,188.17)}, rotate = 319.18] [fill={rgb, 255:red, 74; green, 144; blue, 226 }  ,fill opacity=1 ][line width=0.08]  [draw opacity=0] (8.93,-4.29) -- (0,0) -- (8.93,4.29) -- cycle    ;
\draw  [fill={rgb, 255:red, 0; green, 0; blue, 0 }  ,fill opacity=1 ] (310,414) .. controls (310,411.79) and (311.79,410) .. (314,410) .. controls (316.21,410) and (318,411.79) .. (318,414) .. controls (318,416.21) and (316.21,418) .. (314,418) .. controls (311.79,418) and (310,416.21) .. (310,414) -- cycle ;
\draw  [fill={rgb, 255:red, 0; green, 0; blue, 0 }  ,fill opacity=1 ] (310,374) .. controls (310,371.79) and (311.79,370) .. (314,370) .. controls (316.21,370) and (318,371.79) .. (318,374) .. controls (318,376.21) and (316.21,378) .. (314,378) .. controls (311.79,378) and (310,376.21) .. (310,374) -- cycle ;
\draw  [fill={rgb, 255:red, 0; green, 0; blue, 0 }  ,fill opacity=1 ] (310,334) .. controls (310,331.79) and (311.79,330) .. (314,330) .. controls (316.21,330) and (318,331.79) .. (318,334) .. controls (318,336.21) and (316.21,338) .. (314,338) .. controls (311.79,338) and (310,336.21) .. (310,334) -- cycle ;
\draw  [fill={rgb, 255:red, 0; green, 0; blue, 0 }  ,fill opacity=1 ] (310,294) .. controls (310,291.79) and (311.79,290) .. (314,290) .. controls (316.21,290) and (318,291.79) .. (318,294) .. controls (318,296.21) and (316.21,298) .. (314,298) .. controls (311.79,298) and (310,296.21) .. (310,294) -- cycle ;
\draw  [fill={rgb, 255:red, 0; green, 0; blue, 0 }  ,fill opacity=1 ] (310,254) .. controls (310,251.79) and (311.79,250) .. (314,250) .. controls (316.21,250) and (318,251.79) .. (318,254) .. controls (318,256.21) and (316.21,258) .. (314,258) .. controls (311.79,258) and (310,256.21) .. (310,254) -- cycle ;
\draw  [fill={rgb, 255:red, 0; green, 0; blue, 0 }  ,fill opacity=1 ] (310,214) .. controls (310,211.79) and (311.79,210) .. (314,210) .. controls (316.21,210) and (318,211.79) .. (318,214) .. controls (318,216.21) and (316.21,218) .. (314,218) .. controls (311.79,218) and (310,216.21) .. (310,214) -- cycle ;

\draw  [fill={rgb, 255:red, 0; green, 0; blue, 0 }  ,fill opacity=1 ] (310,174) .. controls (310,171.79) and (311.79,170) .. (314,170) .. controls (316.21,170) and (318,171.79) .. (318,174) .. controls (318,176.21) and (316.21,178) .. (314,178) .. controls (311.79,178) and (310,176.21) .. (310,174) -- cycle ;
\draw  [fill={rgb, 255:red, 0; green, 0; blue, 0 }  ,fill opacity=1 ] (390,254) .. controls (390,251.79) and (391.79,250) .. (394,250) .. controls (396.21,250) and (398,251.79) .. (398,254) .. controls (398,256.21) and (396.21,258) .. (394,258) .. controls (391.79,258) and (390,256.21) .. (390,254) -- cycle ;
\draw  [fill={rgb, 255:red, 0; green, 0; blue, 0 }  ,fill opacity=1 ] (390,214) .. controls (390,211.79) and (391.79,210) .. (394,210) .. controls (396.21,210) and (398,211.79) .. (398,214) .. controls (398,216.21) and (396.21,218) .. (394,218) .. controls (391.79,218) and (390,216.21) .. (390,214) -- cycle ;
\draw  [fill={rgb, 255:red, 0; green, 0; blue, 0 }  ,fill opacity=1 ] (390,174) .. controls (390,171.79) and (391.79,170) .. (394,170) .. controls (396.21,170) and (398,171.79) .. (398,174) .. controls (398,176.21) and (396.21,178) .. (394,178) .. controls (391.79,178) and (390,176.21) .. (390,174) -- cycle ;
\draw  [fill={rgb, 255:red, 0; green, 0; blue, 0 }  ,fill opacity=1 ] (430,174) .. controls (430,171.79) and (431.79,170) .. (434,170) .. controls (436.21,170) and (438,171.79) .. (438,174) .. controls (438,176.21) and (436.21,178) .. (434,178) .. controls (431.79,178) and (430,176.21) .. (430,174) -- cycle ;
\draw [color={rgb, 255:red, 208; green, 2; blue, 27 }  ,draw opacity=1 ][line width=0.75]    (354,334) .. controls (384,301.17) and (343,286.5) .. (354,294) ;
\draw [shift={(365.19,310.12)}, rotate = 455.96] [fill={rgb, 255:red, 208; green, 2; blue, 27 }  ,fill opacity=1 ][line width=0.08]  [draw opacity=0] (8.93,-4.29) -- (0,0) -- (8.93,4.29) -- cycle    ;
\draw [color={rgb, 255:red, 208; green, 2; blue, 27 }  ,draw opacity=1 ][line width=0.75]    (354,295) .. controls (384,262.17) and (343,247.5) .. (354,255) ;
\draw [shift={(365.19,271.12)}, rotate = 455.96] [fill={rgb, 255:red, 208; green, 2; blue, 27 }  ,fill opacity=1 ][line width=0.08]  [draw opacity=0] (8.93,-4.29) -- (0,0) -- (8.93,4.29) -- cycle    ;
\draw [color={rgb, 255:red, 208; green, 2; blue, 27 }  ,draw opacity=1 ][line width=0.75]    (354,254) .. controls (384,221.17) and (343,206.5) .. (354,214) ;
\draw [shift={(365.19,230.12)}, rotate = 455.96] [fill={rgb, 255:red, 208; green, 2; blue, 27 }  ,fill opacity=1 ][line width=0.08]  [draw opacity=0] (8.93,-4.29) -- (0,0) -- (8.93,4.29) -- cycle    ;
\draw [color={rgb, 255:red, 208; green, 2; blue, 27 }  ,draw opacity=1 ][line width=0.75]    (354,214) .. controls (384,181.17) and (343,166.5) .. (354,174) ;
\draw [shift={(365.19,190.12)}, rotate = 455.96] [fill={rgb, 255:red, 208; green, 2; blue, 27 }  ,fill opacity=1 ][line width=0.08]  [draw opacity=0] (8.93,-4.29) -- (0,0) -- (8.93,4.29) -- cycle    ;
\draw  [fill={rgb, 255:red, 0; green, 0; blue, 0 }  ,fill opacity=1 ] (350,214) .. controls (350,211.79) and (351.79,210) .. (354,210) .. controls (356.21,210) and (358,211.79) .. (358,214) .. controls (358,216.21) and (356.21,218) .. (354,218) .. controls (351.79,218) and (350,216.21) .. (350,214) -- cycle ;
\draw  [fill={rgb, 255:red, 0; green, 0; blue, 0 }  ,fill opacity=1 ] (350,174) .. controls (350,171.79) and (351.79,170) .. (354,170) .. controls (356.21,170) and (358,171.79) .. (358,174) .. controls (358,176.21) and (356.21,178) .. (354,178) .. controls (351.79,178) and (350,176.21) .. (350,174) -- cycle ;
\draw  [fill={rgb, 255:red, 0; green, 0; blue, 0 }  ,fill opacity=1 ] (350,254) .. controls (350,251.79) and (351.79,250) .. (354,250) .. controls (356.21,250) and (358,251.79) .. (358,254) .. controls (358,256.21) and (356.21,258) .. (354,258) .. controls (351.79,258) and (350,256.21) .. (350,254) -- cycle ;
\draw  [fill={rgb, 255:red, 0; green, 0; blue, 0 }  ,fill opacity=1 ] (350,294) .. controls (350,291.79) and (351.79,290) .. (354,290) .. controls (356.21,290) and (358,291.79) .. (358,294) .. controls (358,296.21) and (356.21,298) .. (354,298) .. controls (351.79,298) and (350,296.21) .. (350,294) -- cycle ;
\draw  [fill={rgb, 255:red, 0; green, 0; blue, 0 }  ,fill opacity=1 ] (350,334) .. controls (350,331.79) and (351.79,330) .. (354,330) .. controls (356.21,330) and (358,331.79) .. (358,334) .. controls (358,336.21) and (356.21,338) .. (354,338) .. controls (351.79,338) and (350,336.21) .. (350,334) -- cycle ;

\draw (485.38,381) node [anchor=west] [inner sep=0.75pt]    {$\mathcal{A}^{-}$};
\draw (485.38,411) node [anchor=west] [inner sep=0.75pt]    {$\mathcal{A}^{+}$};
\draw (208.4,290) node [anchor=south] [inner sep=0.75pt]  [font=\Large,rotate=-90]  {$\vdots $};
\draw (115,116.6) node [anchor=south] [inner sep=0.75pt]  [font=\Large]  {$\vdots $};
\draw (370,116.6) node [anchor=south] [inner sep=0.75pt]  [font=\Large]  {$\vdots $};
\draw (463.4,290) node [anchor=south] [inner sep=0.75pt]  [font=\Large,rotate=-90]  {$\vdots $};
\draw (33,414) node [anchor=east] [inner sep=0.75pt]  [font=\footnotesize]  {$\nu =0$};
\draw (33,374) node [anchor=east] [inner sep=0.75pt]  [font=\footnotesize]  {$\nu =1$};
\draw (33,334) node [anchor=east] [inner sep=0.75pt]  [font=\footnotesize]  {$\nu =2$};
\draw (33,294) node [anchor=east] [inner sep=0.75pt]  [font=\footnotesize]  {$\nu =3$};
\draw (33,254) node [anchor=east] [inner sep=0.75pt]  [font=\footnotesize]  {$\nu =4$};
\draw (33,214) node [anchor=east] [inner sep=0.75pt]  [font=\footnotesize]  {$\nu =5$};
\draw (33,174) node [anchor=east] [inner sep=0.75pt]  [font=\footnotesize]  {$\nu =6$};
\draw (61,421.4) node [anchor=north west][inner sep=0.75pt]  [font=\scriptsize]  {$\ket{\varphi ^{( 0)}_{0}}$};
\draw (101,341.4) node [anchor=north west][inner sep=0.75pt]  [font=\scriptsize]  {$\ket{\varphi ^{( 2)}_{0}}$};
\draw (141,261.4) node [anchor=north west][inner sep=0.75pt]  [font=\scriptsize]  {$\ket{\varphi ^{( 4)}_{0}}$};
\draw (181,181.4) node [anchor=north west][inner sep=0.75pt]  [font=\scriptsize]  {$\ket{\varphi ^{( 6)}_{0}}$};
\draw (47,122.4) node [anchor=north west][inner sep=0.75pt]  [font=\scriptsize]  {$\ket{\varphi ^{( 0)}_{\nu }}$};
\draw (341,132.4) node [anchor=north west][inner sep=0.75pt]  [font=\scriptsize]  {$\ket{\varphi ^{( 2)}_{\nu -2}}$};
\draw (316,421.4) node [anchor=north west][inner sep=0.75pt]  [font=\scriptsize]  {$\ket{\varphi ^{( 0)}_{0}}$};
\draw (356,341.4) node [anchor=north west][inner sep=0.75pt]  [font=\scriptsize]  {$\ket{\varphi ^{( 2)}_{0}}$};
\draw (396,261.4) node [anchor=north west][inner sep=0.75pt]  [font=\scriptsize]  {$\ket{\varphi ^{( 4)}_{0}}$};
\draw (436,181.4) node [anchor=north west][inner sep=0.75pt]  [font=\scriptsize]  {$\ket{\varphi ^{( 6)}_{0}}$};

\end{tikzpicture}

\caption{The chains of normalised states $\ket{\varphi_\nu^{(2n)}}$. The figure shows the action of $\mathcal{A}^+,\mathcal{A}^-$ on the first two chains of states. $\mathcal{A}^+$ takes a state vertically up in any given column, while $\mathcal{A}^-$ takes a given state to a superposition of states in the row below, the weighting of the superposition of states is different in both examples.}\label{diagram}
\end{figure}

We can represent the space of states defined by $\mathcal{A}^+, \mathcal{A}^-$ diagrammatically, in figure \ref{diagram} each point in the diagram corresponds to a state $\ket{\varphi_\nu^{(2n)}}$ where $n$ indexes the column and $2n+\nu$ indexes the row, with the lowest row and column being defined as the zeroth, i.e. $\ket{\varphi_0^{(0)}}$. The lowest state in each column corresponds to a zero mode \eqref{normgs} and they occur only at even values of the quantum number $\nu$. By defining the chains of states corresponding to each zero mode by successive action of $\mathcal{A}^+$ in lieu of defining that $\mathcal{A}^-$ takes us vertically down a given chain, we find that the action of $\mathcal{A}^-$ takes us to a superposition of states in the row below. This should be read as follows, taking the examples of $\ket{\varphi_3^{(0)}}$ and $\ket{\varphi_1^{(2)}}$ we get
\begin{equation}
\mathcal{A}^-\ket{\varphi_3^{(0)}}=\delta_1\ket{\varphi_2^{(0)}} +\delta_2\ket{\varphi_0^{(2)}}
\end{equation}
and
\begin{equation}
\mathcal{A}^-\ket{\varphi_1^{(2)}}=\delta_3\ket{\varphi_0^{(2)}}+\delta_4\ket{\varphi_2^{(0)}},
\end{equation}
respectively. This is in contrast to the isotropic oscillator where it was found that the generalised ladder operators acted on the same chain analogously to the one-dimensional oscillator \cite{isoand}. We use this basis so we can describe the entire space of states in terms of the generalised ladder operators.

To compute all states in the diagram we make use of a non-commutative binomial theorem \cite{wyss2017noncommutative} to obtain the normal ordered expansion of $(\mathcal{A}^+)^\nu$
\begin{equation}\label{normord}
\begin{split}
 (\alpha b^+ +\beta a^+ b^-)^\nu=\sum_{k=0}^{\floor{\frac{\nu}{2}}}\sum_{j=0}^{\nu-2k} \alpha^{j+k} \beta^{\nu-k-j}  &\frac{\nu!}{(\nu-2k-j)! k! j!2^k}\\
 &\times(b^+)^j  (a^+)^{\nu-k-j}(b^-)^{\nu-2k-j},
\end{split}
\end{equation}
and we define the $\nu$-th unnormalised state in the $n$-th chain by
\begin{equation}
\ket{\psi_\nu^{(2n)}}=(\mathcal{A}^+)^\nu \ket{\varphi_0^{(2n)}}.
\end{equation}
Using \eqref{normord} this yields
\begin{equation}\label{unnorm}
\begin{split}
\ket{\psi_\nu^{(2n)}}=&\frac{1}{\sqrt{\mathcal{N}^{(2n)}_0(\alpha,\beta)}}\sum_{m=0}^n\sum_{k=0}^{\floor{\frac{\nu}{2}}}\sum_{j=\nu-2(k+n-m)}^{\nu-2k}\\
&\Gamma_{m,k,j}^{(2n), (\nu)}(\alpha, \beta)\ket{m+\nu-k-j,2(n-m)-\nu+2k+2j},
\end{split}
\end{equation}
where the coefficients are given by
\begin{equation}
\begin{split}
\Gamma_{m,k,j}^{(2n), (\nu)}(\alpha, \beta)=&\alpha^{j+k} \beta^{\nu-k-j}\gamma_m^{(2n)}(\alpha,\beta) \frac{\nu!}{(\nu-2k-j)! k! j!2^k}\\
&\times\sqrt{\left( m+1\right)_{\nu-k-j}} \sqrt{\left( 2(n-m)-\nu+2k+j+1\right)_{j}}\\
&\times \sqrt{\left( 2(n-m)-\nu+2k+j+1\right)_{\nu-2k-j}}.
\end{split}
\end{equation}
Here $(x)_n=x(x+1)\ldots(x+n-1)$ is the Pochhammer symbol and the functions $\gamma_m^{(2n)}(\alpha,\beta)$ are defined in \eqref{gamcoef}. We mention that the lower limit on the $j$ summation is achieved by excluding states annihilated by $(b^-)^{\nu-2k-j}$. Again it is convenient to normalise the states so we scale \eqref{unnorm} by
\begin{equation}
\ket{\varphi_\nu^{(2n)}}=\frac{1}{\sqrt{\mathcal{N}_{\nu}^{(2n)} (\alpha, \beta)}}\ket{\psi_\nu^{(2n)}},
\end{equation}
where $\mathcal{N}_{\nu}^{(2n)} (\alpha, \beta)= \bra{\psi_\nu^{(2n)}}\ket{\psi_\nu^{(2n)}}$. The function \eqref{condit} is also defined
\begin{equation}
f(\nu)=\frac{\mathcal{N}_\nu^{(2n)} (\alpha, \beta)}{\mathcal{N}_{\nu-1}^{(2n)} (\alpha, \beta)}.
\end{equation}

With every state now defined we are in a position to compute the action of $\mathcal{A}^-$ on an arbitrary state, the states in a given row $\nu$ provide a non-orthogonal basis for the $\nu$-th subspace. Using this basis we can compute
\begin{equation}\label{ami}
\mathcal{A}^- \ket{\varphi_{\nu}^{(2n)}}=\sum_{k=0}^{\floor{\frac{\nu-1}{2}}}\Gamma_k^{(2n)}\ket{\varphi_{\nu-1-2k}^{(2k)}}
\end{equation}
by means of inverting the Gram matrix \cite{Schwerdtfeger:213697}
\begin{equation}\label{gmatrix}
\mathbf{G}=\begin{bmatrix} 
    1 & \bra{\varphi_{\nu-1}^{(0)}}\ket{\varphi_{\nu-3}^{(2)}} & \dots \\
    \vdots & \ddots & \\
    \bra{\varphi_{\nu-1-2\floor{\frac{\nu-1}{2}}}^{(2\floor{\frac{\nu-1}{2}})}}\ket{\varphi_{\nu-1}^{(0)}} &     \dots   & 1
    \end{bmatrix},
\end{equation}
whose matrix elements are $G_{kj}=\bra{\varphi_{\nu+1-2k}^{(2(k-1))}}\ket{\varphi_{\nu+1-2j}^{(2(j-1))}}$ and the diagonal entries $G_{kk}=1$ because the states are normalised. The coefficients \eqref{ami} are then determined by the inversion formula
\begin{equation}
\Gamma_k^{(2n)}=\sum_{j=0}^{\floor{\frac{\nu-1}{2}}}G^{-1}_{kj}\bra{\varphi_{\nu-1-2j}^{(2j)}}\mathcal{A}^-\ket{\varphi_\nu^{(2n)}}.
\end{equation}
In general, for arbitrary $\alpha, \beta$ the off diagonal elements of \eqref{gmatrix} are non-zero but curiously it may be shown that all the zero modes are orthogonal to their corresponding state in the first chain,
\begin{equation}
\bra{\varphi_{0}^{(2\nu)}}\ket{\varphi_{2\nu}^{(0)}}=\frac{\alpha^{2\nu}}{\sqrt{\mathcal{N}_{0}^{(2\nu)} (\alpha, \beta)\mathcal{N}_{2\nu}^{(0)} (\alpha, \beta)}}\sum_{k=0}^\nu (-1)^k \binom{\nu}{k}=0.
\end{equation}

\section{Principle states}\label{secprin}
Focusing on the principle set of states generated from the principle zero mode (the ground state) $\ket{\varphi_0^{(0)}}=\ket{0,0}$, we find the $n=0$ limit of \eqref{unnorm} leads to the unnormalised states
\begin{equation}
 \ket{\psi_\nu^{(0)}}=\sum_{k=0}^{\lfloor\frac{\nu}{2}\rfloor} \alpha^{\nu-k} \beta^{k}  \frac{\nu!}{\sqrt{(\nu-2k)!}\sqrt{k!}2^k}\ket{k,\nu-2k}.
\end{equation}
Normalising the states and absorbing a factor of $\sqrt{\nu!}$ into the definition of the action of $\mathcal{A}^+$ we write the normalised chain as
\begin{equation}\label{main1}
 \ket{\varphi_\nu^{(0)}}=\frac{1}{\sqrt{\mathcal{N}_\nu^{(0)} (\alpha, \beta)}}\sum_{k=0}^{\floor{\frac{\nu}{2}}} \alpha^{\nu-k} \beta^{k}   \sqrt{\binom{\nu}{k}_2}\ket{k,\nu-2k}.
\end{equation}
Here we have introduced a modified binomial coefficient
\begin{equation}
 \binom{n}{k}_t =\frac{n!}{k!(n-tk)! t^{2k}},\quad t\in \mathds{Z}^{\geq 0},
\end{equation}
to elucidate the similarity of the states to the two-mode $\mathfrak{su}(2)$ coherent states. Just as the $\mathfrak{su}(2)$ coherent states may be built from generalised ladder operators picking out degenerate states of the isotropic oscillator \cite{isoand}, it is in this sense that we say the states \eqref{main1} generalise the $\mathfrak{su}(2)$ coherent states to the 2:1 oscillator.

The function $f(\nu)$ (with the additional factor of $\nu$) in \eqref{condit} is defined as
\begin{equation}
f(\nu)=\nu \frac{\mathcal{N}_\nu^{(0)} (\alpha, \beta)}{\mathcal{N}_{\nu-1}^{(0)} (\alpha, \beta)},
\end{equation}
and we observe that the normalisation function may be expressed as
\begin{equation}
\mathcal{N}_\nu^{(0)} (\alpha, \beta)=\left(\frac{\abs{\alpha}\abs{\beta}}{2}\right)^\nu \mathcal{H}_\nu\left(\frac{\abs{\alpha}}{\abs{\beta}} \right),
\end{equation}
where $\mathcal{H}_\nu (x)=\sum_{k=0}^{\lfloor\frac{\nu}{2}\rfloor}\frac{\nu!}{(\nu-2k)!k!}\left( 2x\right)^{\nu-2k}$ are the pseudohermite polynomials \cite{noumi1999}. This yields the desired normalisation condition
\begin{equation}
   \bra{\varphi_\mu^{(0)}}\ket{\varphi_\nu^{(0)}}=\delta_{\mu\nu}.
\end{equation}
Although we have insisted on taking $\mathcal{A}^-=(\mathcal{A}^+)^\dagger$ and as such $\mathcal{A}^-\ket{\varphi_\nu^{(0)}} \not \propto \ket{\varphi_{\nu-1}^{(0)}}$, we can still connect the states exclusively in the $n=0$ chain by using the operator $b^-$, such that 
\begin{equation}
b^-\ket{\varphi_\nu^{(0)}}=\alpha\sqrt{\nu}\sqrt{\frac{\mathcal{N}_{\nu-1}^{(0)} (\alpha, \beta)}{\mathcal{N}_\nu^{(0)} (\alpha, \beta)}}\ket{\varphi_{\nu-1}^{(0)}}.
\end{equation}

A key advantage of these states, because they are built out of the whole space of states of the two-dimensional oscillator, is that we can resolve the identity both in the $\nu$-th subspace and over the full Hilbert space by
\begin{equation}\label{subspace}
 \frac{1}{8\pi^2 \nu!}\int_{\mathds{C}^2}\mathrm{d}^2 \alpha \, \mathrm{d}^2 \beta \, \mathcal{N}_\nu (\alpha, \beta) \frac{e^{-\abs{\alpha}-\frac{\abs{\beta}^2}{4}}}{\abs{\alpha}^{\nu+1}}\ket{\varphi_\nu^{(0)}}\bra{\varphi_\nu^{(0)}}=\mathds{I}_{\nu},
\end{equation}
and
\begin{equation}\label{fullspace}
\sum_{\nu=0}^\infty \left( \frac{1}{8\pi^2 \nu!}\int_{\mathds{C}^2}\mathrm{d}^2 \alpha \, \mathrm{d}^2 \beta \, \mathcal{N}_\nu (\alpha, \beta) \frac{e^{-\abs{\alpha}-\frac{\abs{\beta}^2}{4}}}{\abs{\alpha}^{\nu+1}}\ket{\varphi_\nu^{(0)}}\bra{\varphi_\nu^{(0)}}\right)=\mathds{I}_{\mathcal{H}},
\end{equation}
respectively. For details on the derivation of these results see appendix \ref{appendix}.
 
Thus the states $\left\{\ket{\varphi_{\nu}^{(0)}}\right\}$ form a complete family of states for the Hilbert space of the two-dimensional oscillator, they are equipped with ladder operators and they reproduce the Lissajous figures in their spatial distribution, mimicking the behaviour of the classical 2:1 oscillator \cite{Chen_2003}. In this sense they are a good candidate for the generalisation of the $\mathfrak{su}(2)$ coherent states of the two-dimensional isotropic oscillator.
\begin{figure}[H]
    \centering
    \subfloat{{\includegraphics[scale=0.55]{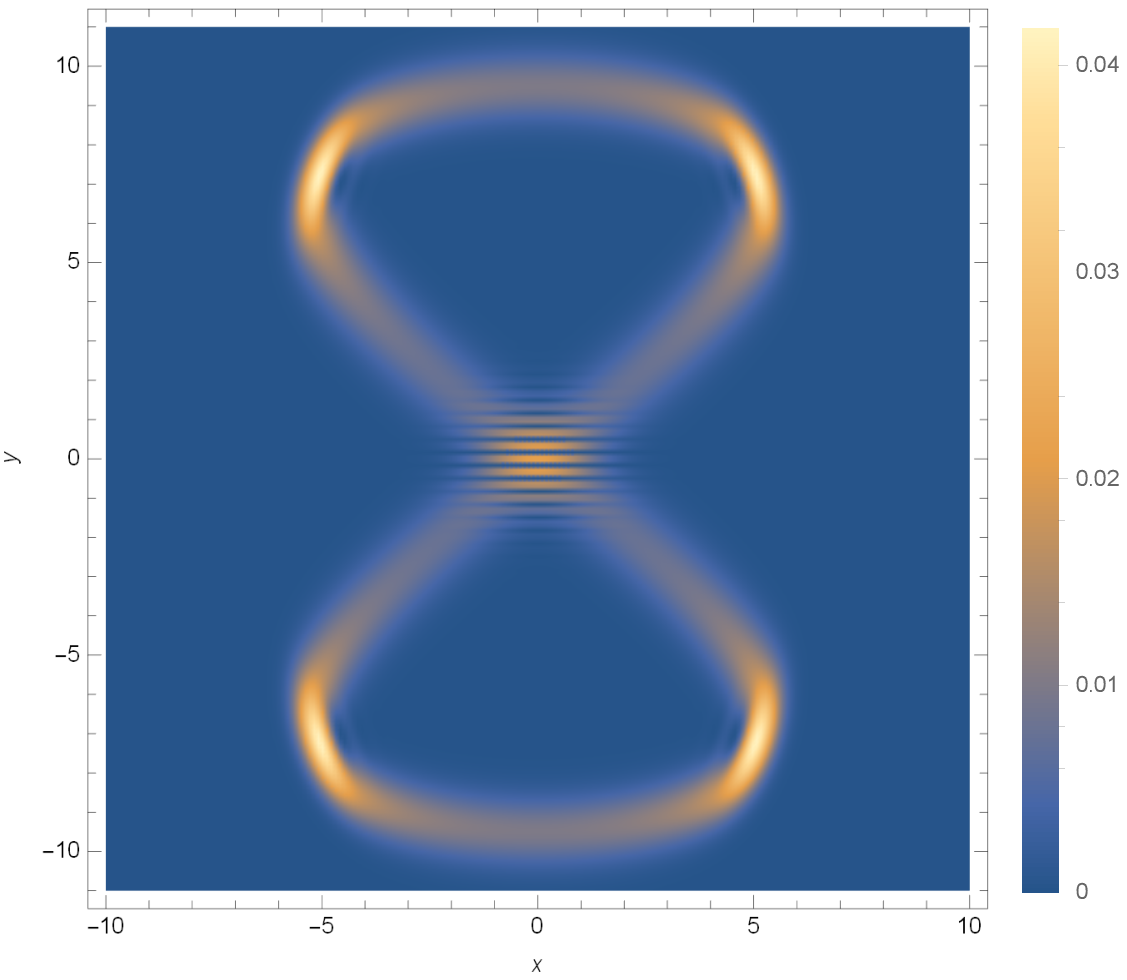} }}%
    \qquad
    \subfloat{{\includegraphics[scale=0.55]{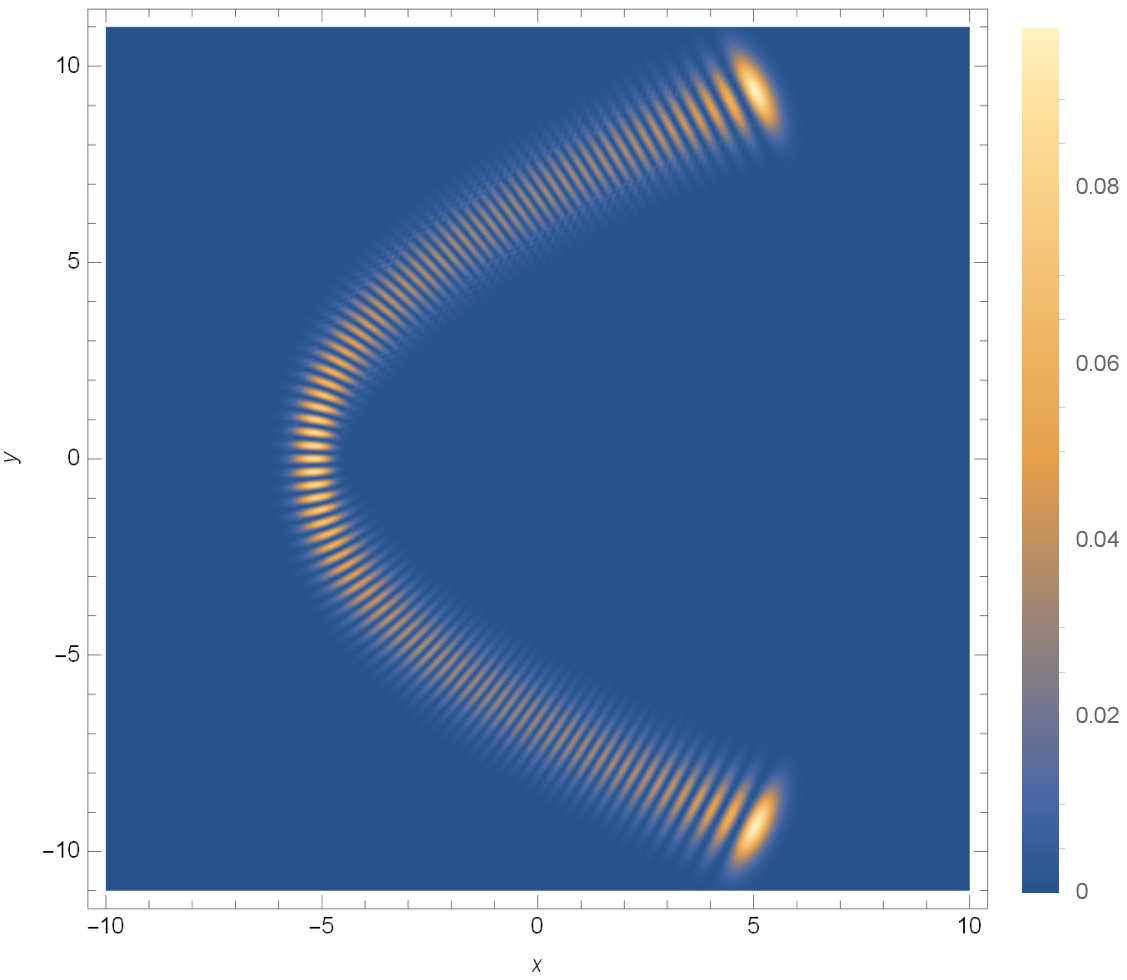}}}%
    \caption{$ \abs{\bra{x,y}\ket{\varphi_{100}^{(0)}}}^2$ with $\alpha=3,\beta=\frac{e^{i\frac{\pi}{2}}}{\sqrt{2}}$ (left) and $\alpha=3,\beta=\frac{1}{\sqrt{2}}$ (right).}%
    \label{fig:example11}%
\end{figure}
Coherent states for harmonic oscillators in their typical presentation follow from three equivalent definitions: eigenstates of the annihilation operator, the action of a unitary displacement operator on the vacuum state, and by a particular infinite superposition of eigenstates \cite{Gazeau:2009zz}. In principle one could construct such displacement operators and eigenstates for the operators presented in \eqref{operators}, however, because their commutation relation is no longer canonical and $\mathcal{A}^-$ does not act correctly as a lowering operator on the same chain of states generated by $(\mathcal{A}^+)^\nu \ket{\varphi_0^{(0)}}$, their generalisation is not straightforward. Additionally because the term $\beta a^+b^-$ in $\mathcal{A}^+$ intrinsically couples the two modes, states generated by the exponential of these operators will not factorise into the product of two one-dimensional harmonic oscillator coherent states. 

We stress that the comparison to be made with the states \eqref{main1} is with the two-mode or Schwinger boson realisation of the $\mathfrak{su}(2)$ coherent states.
\section{Uncertainty relations}\label{uncertainty}
We can calculate the position and momentum uncertainties in the state $\ket{\varphi_\nu^{(0)}}$. In the $a$ mode we get the following product uncertainty relation
\begin{equation}\label{ua}
  \left((\Delta \hat{Q}_a)^2  (\Delta \hat{P}_a)^2\right)_{\ket{\varphi_\nu^{(0)}}}=\frac{1}{4}\left(1+\frac{1}{2}\abs{\alpha}^2\abs{\beta}^2\nu(\nu-1)\frac{\mathcal{N}_{\nu-2}^{(0)}(\alpha, \beta)}{\mathcal{N}_\nu^{(0)} (\alpha, \beta)} \right)^2\geq \frac{1}{4}.
\end{equation}
Similarly for the $b$ mode we find
\begin{equation}\label{ub}
 \left((\Delta \hat{Q}_b)^2(\Delta \hat{P}_b)^2\right)_{\ket{\varphi_\nu^{(0)}}}=\left(\frac{1}{2} +\abs{\alpha}^2 \nu \frac{\mathcal{N}_{\nu-1}^{(0)} (\alpha, \beta)}{\mathcal{N}_\nu^{(0)} (\alpha, \beta)}\right)^2.
\end{equation}
In order to prove these we write $(\Delta \hat{Q}_s)^2  (\Delta \hat{P}_s)^2$ for $s=a,b$ in terms of their ladder operator representations \eqref{ladders} and find that the only non-zero contributions are of the form $\bra{\varphi_\nu^{(0)}} s^+ s^- \ket{\varphi_\nu^{(0)}}$. For $\bra{\varphi_\nu^{(0)}} a^+ a^- \ket{\varphi_\nu^{(0)}}$ we find
\begin{equation}\label{uncertproof}
\begin{split}
 \bra{\varphi_\nu^{(0)}} a^+ a^- \ket{\varphi_\nu^{(0)}} &= \frac{1}{\mathcal{N}_\nu^{(0)} (\alpha, \beta)}\sum_{k=0}^{\lfloor\frac{\nu}{2}\rfloor} \abs{\alpha}^{2(\nu-k)} \abs{\beta}^{2k} \frac{k}{(\nu-2k)!k!2^{2k}}\\
 &= \frac{1}{\mathcal{N}_\nu^{(0)} (\alpha, \beta)}\sum_{k=0}^{\lfloor\frac{\nu}{2}\rfloor-1} \abs{\alpha}^{2(\nu-1-k)} \abs{\beta}^{2(k+1)} \frac{1}{(\nu-2-2k)!k!2^{2(k+1)}}\\
 &=\frac{1}{4}\abs{\alpha}^2\abs{\beta}^2\nu(\nu-1)\frac{\mathcal{N}_{\nu-2}^{(0)}(\alpha, \beta)}{\mathcal{N}_\nu^{(0)} (\alpha, \beta)},
\end{split}
\end{equation}
where in the second line of \eqref{uncertproof} we changed the summation index $k\rightarrow k+1$ and in the third line we used the fact that $\floor*{\frac{\nu}{2}}-1=\floor*{\frac{\nu-2}{2}}$ to rewrite the summation in terms of the normalisation function with index $\nu-2$. The same principle is used to compute terms of the form $\bra{\varphi_\nu^{(0)}} b^+ b^- \ket{\varphi_\nu^{(0)}}$ allowing us to recover the uncertainty relations in \eqref{ua} and \eqref{ub}.
\begin{figure}[H]
    \centering
    \subfloat[]{{\includegraphics[scale=0.65]{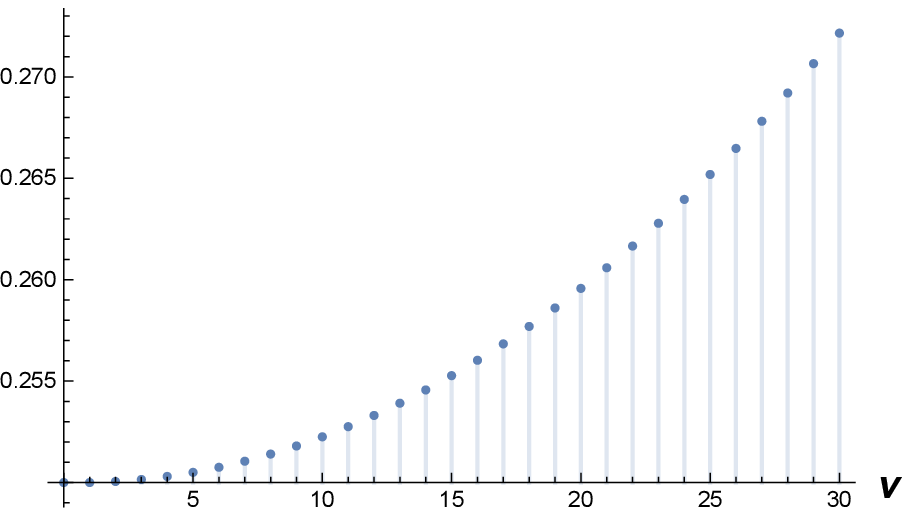} }}%
    \qquad
    \subfloat[]{{\includegraphics[scale=0.65]{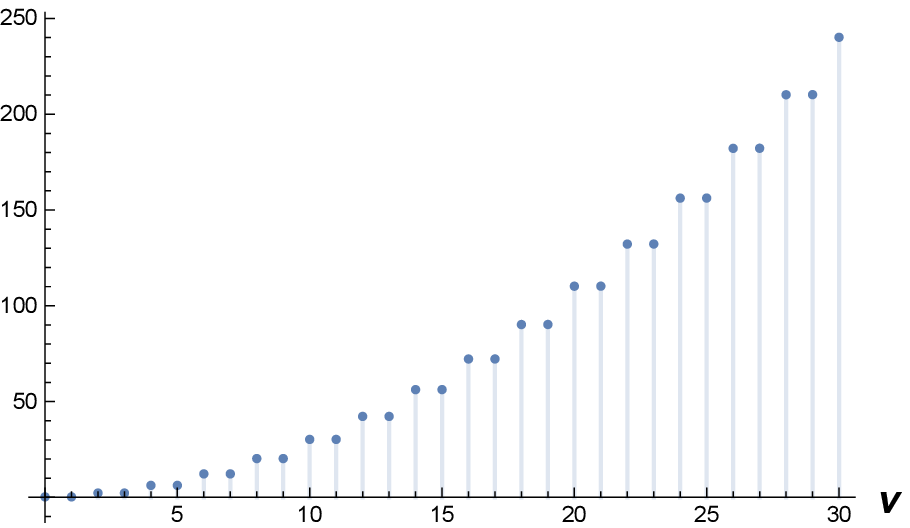}}}\\
    \subfloat[]{{\includegraphics[scale=0.65]{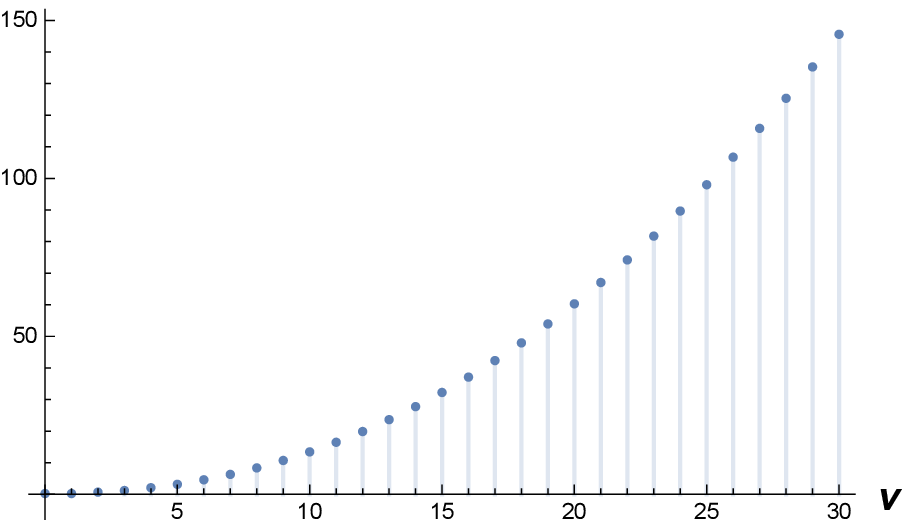}}}%
    \caption{$ \left((\Delta \hat{Q}_a)^2  (\Delta \hat{P}_a)^2\right)_{\ket{\varphi_\nu^{(0)}}}$ as a function of $\nu$. (a) $\alpha=100,\beta=1$, (b) $\alpha=1,\beta=100$, (c) $\alpha=1,\beta=1$.}%
    \label{amod}%
\end{figure}
\begin{figure}[H]
    \centering
    \subfloat[]{{\includegraphics[scale=0.65]{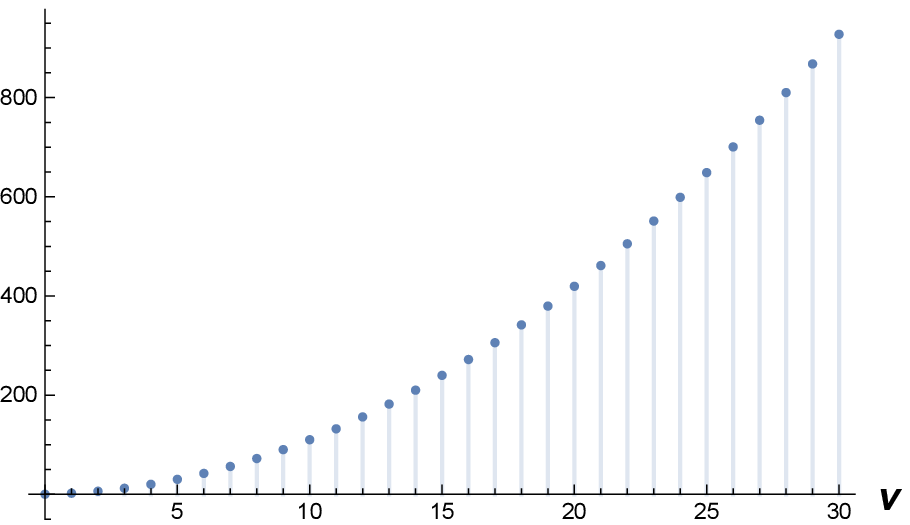} }}%
    \qquad
    \subfloat[]{{\includegraphics[scale=0.65]{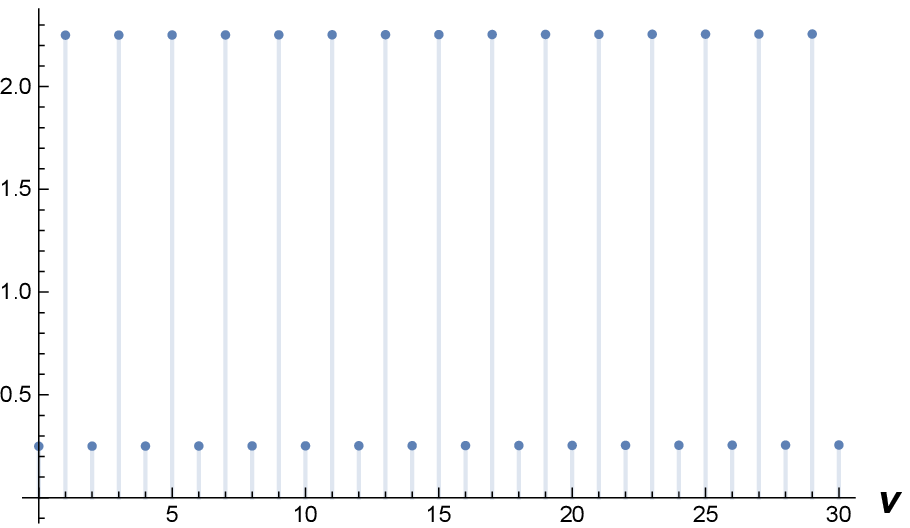}}}\\
    \subfloat[]{{\includegraphics[scale=0.65]{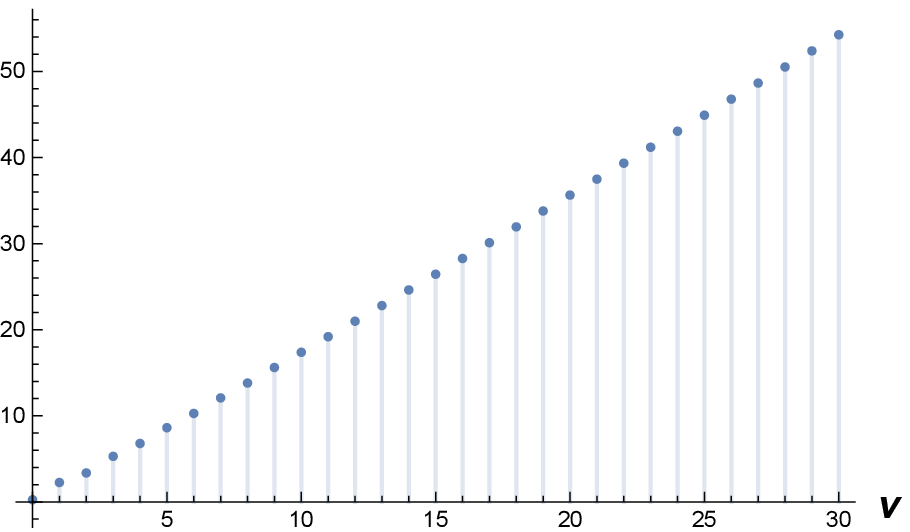}}}%
    \caption{$ \left((\Delta \hat{Q}_b)^2  (\Delta \hat{P}_b)^2\right)_{\ket{\varphi_\nu^{(0)}}}$ as a function of $\nu$. (a) $\alpha=100,\beta=1$, (b) $\alpha=1,\beta=100$, (c) $\alpha=1,\beta=1$.}%
    \label{bmod}%
\end{figure}

The states satisfy the physical condition on the position-momentum uncertainty relation in both the $a$ and $b$ modes, $(\Delta \hat{Q}_i)^2  (\Delta \hat{P}_i)^2\geq \frac{1}{4}$. In (a) of figure \ref{amod} we have approximated the limit $\abs{\alpha} \gg \abs{\beta}$, so the $b$ mode is the dominant mode. In this limit the mixing of the modes is such that the effects of the $a$ mode are diluted by that of the $b$ mode and as a result we see that the product of uncertainties in the $a$ mode increase relatively slowly when compared with (a) of figure \ref{bmod}, where the uncertainty relation increases much more rapidly as a function of $\nu$.

In (b) of figures \ref{amod} and \ref{bmod} we approximate the limit $\abs{\beta} \gg \abs{\alpha}$, in this case the mixing term in $\mathcal{A}^+$ is dominant and as such we see a staggering pattern in the uncertainties of the $a$ mode, for an even value of $\nu=2n$, we observe that the uncertainty in the state $2n+1$ is approximately equal. Meanwhile in the $b$ mode the effect is such that for odd values of $\nu$ the product of uncertainties $ \left((\Delta \hat{Q}_b)^2  (\Delta \hat{P}_b)^2\right)_{\ket{\varphi_{2\nu+1}^{(0)}}}\sim 2.25$ while for even values $ \left((\Delta \hat{Q}_b)^2  (\Delta \hat{P}_b)^2\right)_{\ket{\varphi_{2\nu}^{(0)}}}\sim 0.25$, this is due to the the fact that the state $\ket{\varphi_{2\nu+1}^{(0)}}$ contains the same number of basis states $\ket{k,\nu-2k}$ as the state $\ket{\varphi_{2\nu}^{(0)}}$ but with a larger number in the $b$ mode and therefore there is a larger uncertainty associated to odd values of $\nu$. The staggering in the $a$ mode is a result of the anisotropy of the system, because $a^+$ is worth two quanta (relative to $b^+$ being worth one quantum), we only see a measurable increase in its uncertainty when $\nu$ increases by two.

In (c) of figures \ref{amod} and \ref{bmod} we choose $\alpha=\beta$, so that the mixing term has the same weight as the term adding only $b$ modes, we find that the uncertainty in the $a$ mode is parabolic in $\nu$ while the uncertainty in $b$ is linear.

\section{Conclusion}\label{conc}
In this article we have presented a new set of states for the 2:1 quantum anisotropic oscillator. They admit a ladder operator construction and a resolution of the identity, moreover, they reproduce the Lissajous figures of \cite{Chen_2003} and generalise the $\mathfrak{su}(2)$ coherent states of the two dimensional isotropic oscillator. Furthermore we find that the mixing of the modes means that the uncertainty relations for each mode are codependent, and, for certain choices of parameters there are interesting staggering patterns on the respective uncertainties.

We also found in the general construction that we can build chains of states from higher energy zero modes defined by the operator $\mathcal{A}^-$, and as such building states by defining ladder operators rather than defining expansion coefficients leads to a rich structure in the space of states. We focused on the principle chain of states in this paper, but similar analyses may be completed for the other chains of states.

It would be interesting to consider other types of ladder operators, it is clear that $\mathcal{A}^+$ is not a unique choice which picks out the appropriate energy eigenstates, though it appears to be the simplest one. Other choices of ladder operator will be, in general, harder to analytically normal order, though the solution to this may lie in extending some results in \cite{doi:10.1119/1.2723799} to the ordering of multidimensional operators. As well, the inclusion of accidentally degenerate states into this formalism will require additional caution because they cannot be predicted by symmetry arguments and therefore the process of defining ladder operators to capture this may need some modification. Finally, applying these techniques to multidimensional systems other than the harmonic oscillator would be interesting, a system such as the 2D Morse potential which is a more realistic modelling of the vibrations of molecules \cite{PhysRevA.76.052114}, or the Pais-Uhlenbeck oscillator \cite{PhysRev.79.145} which has been studied as one possible path to a theory of quantum gravity. The techniques presented in this paper may lead to interesting classes of states for these systems.

\section{Acknowledgements}
J. Moran acknowledges the support of the D\'epartement de physique at the Universit\'e de Montr\'eal. V. Hussin acknowledges the support of research grants from NSERC of Canada. I. Marquette was supported by Australian Research Council Future Fellowship FT180100099.
\begin{appendices}
\section{Resolution of the identity}\label{appendix}
To compute the resolution of the identity we write the parameters $\alpha=\abs{\alpha}e^{i\theta}, \beta=\abs{\beta}e^{i\phi}$, in polar form and considering for some measure $\mu_\nu\left(\abs{\alpha}, \abs{\beta}\right)$,
\begin{equation}
\int_{0}^\infty\mathrm{d} \abs{\alpha} \, \abs{\alpha} \int_0^\infty \mathrm{d} \abs{\beta} \, \abs{\beta} \int_0^{2\pi} \mathrm{d}\theta  \int_0^{2\pi} \mathrm{d} \phi \, \mu_\nu\left(\abs{\alpha}, \abs{\beta}\right)\mathcal{N}_\nu (\alpha, \beta)\ket{\varphi_\nu^{(0)}}\bra{\varphi_\nu^{(0)}}.
\end{equation}
The angular integrations over $\theta, \phi$, yield a factor of $4\pi^2$ multiplied by a Kronecker delta matching the summation indices of $\ket{\varphi_\nu^{(0)}}\bra{\varphi_\nu^{(0)}}$. To address the radial integrations
\begin{equation}\label{denominator}
\begin{split}
&4\pi^2 \nu! \int_{0}^\infty\mathrm{d} \abs{\alpha} \int_0^\infty \mathrm{d} \abs{\beta}\, \mu_\nu\left(\abs{\alpha}, \abs{\beta}\right)\\
 &\times\sum_{k=0}^{\lfloor\frac{\nu}{2}\rfloor} \abs{\alpha}^{2(\nu-k)+1} \abs{\beta}^{2k+1} \frac{1}{(\nu-2k)!k!2^{2k}}\ket{k,\nu-2k}\bra{k,\nu-2k},
\end{split}
\end{equation}
we make use of the following identities
\begin{equation}
 \int_0^\infty \mathrm{d} x\, x^{2k+1} e^{-c x^2} = \frac{k!}{2(c^{k+1})}, \quad k\in\mathds{Z}, c>0,
\end{equation}
and
\begin{equation}
 \int_0^\infty \mathrm{d} x\, x^{n} e^{-d x} = \frac{n!}{d^{n+1}}, \quad n\in \mathds{Z}^{\geq0}, \textrm{Re}(d) >0,
\end{equation}
after which we observe that choosing the measure
\begin{equation}
\mu_\nu\left(\abs{\alpha}, \abs{\beta}\right)=  \frac{1}{8\pi^2 \nu!} \frac{e^{-\abs{\alpha}-\frac{\abs{\beta}^2}{4}}}{\abs{\alpha}^{\nu+1}},
\end{equation}
produces the correct factorial terms to cancel the denominator of \eqref{denominator} and we obtain 
\begin{equation}\label{evenodd}
 \sum_{k=0}^{\lfloor\frac{\nu}{2}\rfloor}\ket{k,\nu-2k}\bra{k,\nu-2k} \equiv \mathds{I}_\nu,
\end{equation}
in agreement with \eqref{subspace}. 

To obtain the identity operator on the full Hilbert space we sum over each partition $\nu$ using the reverse of the Cauchy product formula
\begin{equation}
\left( \sum_{n=0}^\infty x_n\right) \left( \sum_{m=0}^\infty y_m\right) =\sum_{k=0}^\infty \sum_{l=0}^k x_l y_{k-l}.
\end{equation}
We consider $\nu$ even and odd separately in \eqref{evenodd}. For $\nu \rightarrow 2\nu'$ even
\begin{equation}\label{even1}
\sum_{\nu'=0}^\infty \sum_{k=0}^{\nu'}\ket{k,2\nu'-2k}\bra{k,2\nu'-2k}  = \sum_{n=0}^\infty \sum_{m=0}^\infty \ket{n,2m},\bra{n,2m},
\end{equation}
and for $\nu \rightarrow 2\nu'+1$ odd 
\begin{equation}\label{odd1}
\sum_{\nu'=0}^\infty \sum_{k=0}^{\nu'}\ket{k,2\nu'-2k+1}\bra{k,2\nu'-2k+1}  = \sum_{n=0}^\infty \sum_{m=0}^\infty \ket{n,2m+1}\bra{n,2m+1}.
\end{equation}
The combination of \eqref{even1} and \eqref{odd1} gives the desired result \eqref{fullspace}
\begin{equation}
\sum_{\nu=0}^\infty\mathds{I}_\nu  =\sum_{n=0}^\infty \sum_{m=0}^\infty \ket{n,m}\bra{n,m}=\mathds{I}_\mathcal{H}.
\end{equation}
\end{appendices}
\bibliographystyle{aip}
\bibliography{refs.bib}

\end{document}